\documentclass[10pt,journal]{IEEEtran}

\usepackage{amssymb}
\usepackage{amsmath}
\usepackage{mathrsfs }
\usepackage{stmaryrd}
\usepackage{amsthm}
\usepackage[noadjust]{cite}
\usepackage{graphicx}
\usepackage{subfigure}
\graphicspath{{pdfs/},{images/}}
\usepackage{xcolor}
\usepackage{colortbl,dcolumn}
\usepackage{algorithm}
\usepackage{algorithmic}
\usepackage{diagbox}
\usepackage{booktabs}
\allowdisplaybreaks[4]
\usepackage{stfloats}
\SetSymbolFont{stmry}{bold}{U}{stmry}{m}{n}
\newtheorem{definition}{Definition}
\newtheorem{lemma}{Lemma}
\newtheorem{theorem}{Theorem}
\newtheorem{proposition}{Proposition}
\newtheorem{corollary}{Corollary}
\newtheorem{remark}{Remark}
\newtheorem{problem}{Problem}

\newcommand{\defref}[1]{Definition~\ref{#1}}
\newcommand{\lemref}[1]{Lemma~\ref{#1}}
\newcommand{\thmref}[1]{Theorem~\ref{#1}}
\newcommand{\propref}[1]{Proposition~\ref{#1}}
\newcommand{\corref}[1]{Corollary~\ref{#1}}
\newcommand{\figref}[1]{Fig.~\ref{#1}}
\newcommand{\tabref}[1]{Table~\ref{#1}}
\newcommand{\secref}[1]{Section~\ref{#1}}
\newcommand{\apxref}[1]{Appendix~\ref{#1}}
\newcommand{\probref}[1]{Problem~\ref{#1}}

\newcommand{\rekref}[1]{Remark~\ref{#1}}

\begin{document}

\title{Formation Under Communication Constraints: Control Performance Meets Channel Capacity}

\author{Yaru Chen, Yirui Cong, \IEEEmembership{Member, IEEE}, Xiangyun Zhou, \IEEEmembership{Fellow, IEEE}, Long Cheng, \IEEEmembership{Fellow, IEEE},

and Xiangke Wang, \IEEEmembership{Senior Member, IEEE}
\thanks{Yaru Chen, Yirui Cong, and Xiangke Wang are with the College of Intelligence Science and Technology, National University of Defense Technology, Changsha 410073, China (e-mail: chenyaru21@nudt.edu.cn; congyirui11@nudt.edu.cn; xkwang@nudt.edu.cn).}
\thanks{Xiangyun Zhou is with the School of Engineering, the Australian National University, Canberra, ACT 2601, Australia (e-mail: xiangyun.zhou@anu.edu.cn).}
\thanks{ Long Cheng is with the State Key Laboratory of Multimodal Artificial Intelligence Systems, Institute of Automation, Chinese Academy of Sciences, Beijing 100190, China, and also with the School of Artificial Intelligence, University of Chinese Academy of Sciences, Beijing 100049, China (e-mail: long.cheng@ia.ac.cn).}}

\IEEEtitleabstractindextext{
\begin{abstract}

In wireless communication-based formation control systems, the control performance is significantly impacted by the channel capacity of each communication link between agents. This relationship, however, remains under-investigated in the existing studies. To address this gap, the formation control problem of classical second-order multi-agent systems with bounded process noises was considered taking into account the channel capacity. More specifically, the model of communication links between agents is first established, based on a new concept -- guaranteed communication region, which characterizes all possible locations for successful message decoding in the present of control-system uncertainty. Furthermore, we rigorously prove that, the guaranteed communication region does not unboundedly increase with the transmission time, which indicates an important trade-off between the guaranteed communication region and the data rate. The fundamental limits of data rate for any desired accuracy are also obtained. Finally, the integrated design to achieve the desired formation accuracy is proposed, where an estimation-based controller and transmit power control strategy are developed.

\end{abstract}

}

\maketitle

\IEEEdisplaynontitleabstractindextext

\IEEEpeerreviewmaketitle

\section{Introduction}\label{sec:Introduction}

\subsection{Motivation and Related Work}\label{sec:Motivation and Related Work}

Formation control of multi-agent systems (MASs) has attracted considerable interest over the last decades due to its wide applications in many regions~\cite{OH2015424,NOWZARI20191}.
The performance of formation control is fundamentally limited by the quality of communication links between agents.
Since the agents usually exchange information through wireless networks, the links are constrained by the physical layer of wireless communications.
However, the literature lacks a comprehensive understanding of such communication links in the formation control system.

In control theory, the model of communication links can be classified into two main classes:
\begin{itemize}
    \item \textbf{Location-independent class:}
    The distance information between agents is not explicitly reflected in modeling communication links.
    For consensus-based formation control, each communication link is associated with a non-zero number, representing the strength of this link in regard to the control protocol, which is usually one (unweighted)~\cite{Zong2019,Zheng2019} or a positive number (weighted)~\cite{Network_Topology_and_Communication_Data_Rate_for_Consensusability_of_Discrete_time_Multi_Agent_Systems,Ma2023,Liu2024,Lin2024}.
    For distance-based formation control~\cite{Sun2017}, each communication link specifies the distance constraint of the target formation.
    For bearing-based formation control, such as~\cite{Zhao2016}, each communication link is associated with a constant orthogonal projection matrix, which geometrically projects any vector onto the orthogonal complement of itself, representing the bearing constraint of the target formation.
    For complex Laplacian-based formation control~\cite{Lin2014}, each communication link is related to a non-zero complex number, and the complex weight is designed appropriately to satisfy the linear formation configuration constraints.
    Similarly, for affine formation~\cite{Zhao2018}, the weight assigned to each communication link is defined by an equilibrium stress, which can be a positive number (attracting force), or a negative number (repelling force).
    
    \item \textbf{Location-dependent class:} Different from the location-independent class, the model of communication links in this class utilizes the distance information between agents, through a map from inter-agent distance to edge weight, e.g., the interaction function~\cite{Second-order_consensus_of_multi-agent_systems_under_limited_interaction_ranges}.
    In~\cite{Circle_Formation_Control_of_Mobile_Agents_With_Limited_Interaction_Range,Santilli2022}, the interaction function is discontinuous, and the edge weight equals one when the corresponding inter-agent distance is less than a certain value, called the interaction radius (in which case the communication link exists); otherwise, the weight is zero.
    In~\cite{Preserving_Strong_Connectivity_in_Directed_Proximity_Graphs}, the interaction function was represented by a smooth non-increasing bump function, and thus the weight varies continuously with the inter-agent distance.
    Note that the interaction radii in the above studies are time-invariant.
    In~\cite{Decentralized_Estimation_and_Control_for_Preserving_the_Strong_Connectivity_of_Directed_Graphs}, the interaction radius was designed to be time-varying for preserving strong connectivity.
    
\end{itemize}

Indeed, the distance is an important factor affecting the model of communication links.
However, it cannot fully reveal the essential constraints from the physical layer.
Even though many issues of control systems have been well tackled with the simplified model of communication links in~\cite{Zong2019}-\cite{Decentralized_Estimation_and_Control_for_Preserving_the_Strong_Connectivity_of_Directed_Graphs}, a more realistic communication model is still necessary to achieve more efficient collaborative control for MASs.

In communication theory, communication links are often characterized by the relationship between the data rate and the channel capacity (e.g., Shannon capacity~\cite{A_mathematical_theory_of_communication}), which depends on not only the locations but also the bandwidth, transmit power, noise, etc.,~\cite{Rappaport1996}.
More specifically, reliable communication is only possible when the data rate is below the channel capacity.
To describe the communication performance between mobile agents/nodes, stochastic geometry~\cite{HaenggiM2012_BOOK} was usually employed to model the classical mobilities in wireless networks;
for general mobilities,~\cite{7076612} proposed a compound Gaussian point process to describe/approximate arbitrary real-time communication property including the channel capacity.
Without relying on mobility models,~\cite{8345703} investigated the channel capacity between two agents (unmanned aerial vehicles, UAVs) based on random trajectories.
Unlike~\cite{HaenggiM2012_BOOK, 7076612, 8345703} without explicitly considering the control inputs,~\cite{9162475} analyzed the expected channel capacity when the formation control system achieved the second-moment stability, which has a non-vanishing lower bound;~\cite{Fan2024} developed two movement control strategies for multiple UAV systems regarding the static and mobile user equipments, and then derived the analytical expression of the ergodic rate for a typical user equipment to evaluate the transmission performance.
Even though the communication performance was analyzed in~\cite{HaenggiM2012_BOOK}-\cite{Fan2024} (particularly, in~\cite{9162475} and~\cite{Fan2024}, with controlled movements), the effects of the control performance on the channel capacity were not studied.

In formation control systems relying on wireless links, the data rate is closely related to the performance of formation control.
For instance, one can reduce the data rate (at the expense of the formation accuracy) to satisfy the channel capacity constraint, which increases the number of reliable communication links (to enhance the formation stability).
As a result, the control performance and the channel capacity are intriguingly coupled, through the communication links.
This fact necessitates a joint study of control and communication theories to better understand the relationship between control performance and channel capacity.

\subsection{Our Contributions}\label{sec:Our Contributions}

In this article, the formation control problem of classical second-order MASs with bounded process noises, each agent using a unique communication band\footnote{This setting can avoid communication interferences and make us concentrate on the channel capacity itself.
This is because the capacity region for multi-user networks is largely an open research problem~\cite{ElGamalA2011BOOK};
even regarding interference as noise, we can hardly derive a closed-form throughput region~\cite{7740102}.
For scenarios with multiple communicating pairs, the frequency division multiple access protocol is commonly adopted to guarantee different communication nodes are allocated with unique bands~\cite{Chen2023,Myung2006}.},
is considered to better understand control performance versus channel capacity.
In particular, we focus on analyzing the fundamental limits of data rate\footnote{The fundamental limits reflect the intrinsic correlation between control performance and channel capacity.
In this work, the round-off error of the state information is assumed to be neglectable;
this means we did not take the source coding scheme into account, whose fundamental limit remains unknown in the information-theoretic sense~\cite{LiT2012,QiuZ2016}.
}
under a given control accuracy level and providing an integrated design of control and communication.
The main contributions are as follows:

\begin{itemize}
    \item   From the joint perspective of control and communication, we first establish the model of communication links between agents, based on a new concept -- guaranteed communication region.
    The guaranteed communication region of a transmitter agent characterizes all possible locations for successful message decoding (i.e., satisfying constraints of channel capacities).
    It utilizes the information from the data rate, communication bandwidth, transmit power, power of noise, and location uncertainty of the transmitter agent.

    \item   More importantly, we analyze the fundamental limits of data rate under any desired formation accuracy, where the guaranteed communication region plays the key role.
    We rigorously prove the property regarding the radius of the guaranteed communication region.
    Surprisingly, it is concave (with a maximum) with respect to (w.r.t.) the transmission time.
    This implies a fundamental trade-off between the guaranteed communication region and the data rate, which is different from the traditional communication theory that decreasing the data rate will generally increase the communication region.

    \item   Finally, we propose an integrated design of control and communication for the MAS with the desired data rate and formation accuracy.
    Specifically, for the control subsystem, we develop an estimation-based controller to guarantee the formation stability under the transmission model determined by the communication subsystem;
    for the communication subsystem, we provide a distributed transmit power control strategy (which effectively changes the channel capacities) to maintain the required communication links by the control subsystem.
    The integrated design fully utilizes the limited communication resources and can handle the time-varying communication conditions, which significantly improves the ability of cooperative control for MASs constrained by the communication physical layer.
    
\end{itemize}

\subsection{Article Organization}\label{sec:Article Organization}

The remainder of the article is organized as follows.
In \secref{sec:System Model and Problem Description}, we provide the system model and problem description.
\secref{sec:Analysis of Fundamental Limit for Data Rate} analyzes the fundamental limits of the data rate, which has three subsections:
in \secref{sec:Analysis of Uncertain Position Range}, the position range is given to describe the positional uncertainties of agents;
in \secref{sec:Property of Guaranteed Communication Radius}, an important property of the guaranteed communication radius is established;
in \secref{sec:Fundamental Limit for Data Rate}, the fundamental limits of the data rate are characterized.
Then, we propose an integrated design of control and communication in \secref{sec:Closed-Loop Design of Formation Control and Communication Transmission}.
Simulation results are provided in \secref{sec:Simulation} to illustrate the effectiveness of our design.
Finally, concluding remarks are made in \secref{sec:Conclusion And Future Work}.

\subsection{Notation}\label{sec:Notation}

Throughout the article, $\mathbb{Z}_+$, $\mathbb{N}_0$, and $\mathbb{R}^+$ denote the sets of positive integers, nonnegative integers, and positive real numbers, respectively.
$\mathbb{R}^n$ stands for the $n$-dimensional Euclidean space, and $\mathbb{R}^{n \times n}$ denotes the set of all $n \times n$ real matrices.
$\mathbf{0}_n \in \mathbb{R}^n$ and $\mathbf{1}_n \in \mathbb{R}^n$ are the column vectors with each entry equal to $0$ and $1$; $\mathbf{0}_{n \times m}$ and $\mathbf{1}_{n \times m}$ are the $n \times m$ matrices with each entry equal to~0 and~1; $I_n \in \mathbb{R}^{n \times n}$ denotes an $n$-dimensional identity matrix.
If not specified, $\|\cdot\|$ refers to the Euclidean norm or the corresponding induced matrix norm.
Given two sets $\mathcal{S}_1$ and $\mathcal{S}_2$ in a Euclidean space, the Minkowski sum of $\mathcal{S}_1$ and $\mathcal{S}_2$ is $\mathcal{S}_1 \oplus \mathcal{S}_2=\left\lbrace s_1+s_2 \colon s_1 \in \mathcal{S}_1, s_2 \in \mathcal{S}_2 \right\rbrace $.
We use $\mathbf{x}$, $x$, and $\llbracket\mathbf{x}\rrbracket$ to denote an uncertain variable\footnote{Consider a sample space $\Omega$. A measurable function $\mathbf{x}: \Omega \to \mathcal{X}$ from the sample space $\Omega$ to the measurable set $\mathcal{X}$ is called an uncertain variable~\cite{A_Nonstochastic_Information_Theory_for_Communication_and_State_Estimation}.
The realization and range of an uncertain variable (say $\mathbf{x}$) can be defined as $\mathbf{x}(\omega): = x$ and $\llbracket\mathbf{x}\rrbracket: = \left\{\mathbf{x}(\omega)\colon \omega \in \Omega\right\}$, respectively.}, its realization, and its range, respectively.
For vector $x \in \mathbb{R}^{n}$, the $n$-dimensional closed ball and open ball can be denoted by $\mathcal{B}[c,r]:=\{ x \colon \|x-c\| \leq r \}$ and ${\mathcal{B}}(c,r):=\{ x \colon \|x-c\| < r \}$, where $c \in \mathbb{R}^n$ and $r \in \mathbb{R}$ represent the center and the radius, respectively.
Denote $S(\Psi)$ as the set of solutions to $\Psi \geq 0$.
$S^{\min}(\Psi)$ and $S^{\max}(\Psi)$ represent the minimum and maximum positive integers in~$S(\Psi)$.

\section{System Model and Problem Description}\label{sec:System Model and Problem Description}
\subsection{System Model}\label{sec:System Model}

Consider $N$ agents moving in an $n$-dimensional space and sharing information through wireless communications. 
The MASs are affected by bounded process noises which can be denoted by uncertain variables.
The $i$\textsuperscript{th} ($i \in \mathcal{V}:=\{1,\ldots,N\}$) agent is governed by the following discrete-time dynamics
\begin{equation}\label{eqn:Second-Order of Position and Velocity}
\begin{cases}
    \mathbf{p}_{i,k+1} = \mathbf{p}_{i,k}+h\mathbf{v}_{i,k}+\frac{h^2}{2}\mathbf{u}_{i,k}+\mathbf{w}_{i,k}^{p}, \\
    \mathbf{v}_{i,k+1} = \mathbf{v}_{i,k}+h\mathbf{u}_{i,k}+\mathbf{w}_{i,k}^{v},
\end{cases}
\end{equation}
where $h$ and $t = kh$ ($k \in \mathbb{N}_0$) are the sampling period and sampling instant, respectively.
In \eqref{eqn:Second-Order of Position and Velocity}, $\mathbf{p}_{i,k}$, $\mathbf{v}_{i,k}$, $\mathbf{u}_{i,k}$, $\mathbf{w}_{i,k}^{p}$, and $\mathbf{w}_{i,k}^{v}$ are the position, velocity, control input, position process noise, and velocity process noise of agent $i$ with their realizations $p_{i,k} \in \llbracket\mathbf{p}_{i,k}\rrbracket \subseteq \mathbb{R}^n$, $v_{i,k} \in \llbracket\mathbf{v}_{i,k}\rrbracket \subseteq \mathbb{R}^n$, $u_{i,k} \in \llbracket\mathbf{u}_{i,k}\rrbracket \subseteq \mathbb{R}^n$, $w_{i,k}^{p} \in \llbracket\mathbf{w}_{i,k}^{p}\rrbracket \subseteq \mathbb{R}^n$, and $w_{i,k}^{v} \in \llbracket\mathbf{w}_{i,k}^{v}\rrbracket \subseteq \mathbb{R}^n$, respectively.
The uncertain variables $\mathbf{p}_{i,k}$, $\mathbf{v}_{i,k}$, $\mathbf{u}_{i,k}$, $\mathbf{w}_{i,k}^{p}$, and $\mathbf{w}_{i,k}^{v}$ are (mutually) unrelated~\cite{A_Nonstochastic_Information_Theory_for_Communication_and_State_Estimation}.
The ranges of $\mathbf{w}_{i,k}^p$ and $\mathbf{w}_{i,k}^v$ are $n$-dimensional closed balls $\llbracket\mathbf{w}_{i,k}^{p}\rrbracket=\mathcal{B}[\mathbf{0}_n,r_{w_p}(i)]$ and $\llbracket\mathbf{w}_{i,k}^v\rrbracket=\mathcal{B}[\mathbf{0}_n,r_{w_v}(i)]$, where $r_{w_p}(i),~r_{w_v}(i) \in \mathbb{R}$ are constants.
The initial position $p_{i,0}$ and velocity $v_{i,0}$ for $i \in \mathcal{V}$ are determinate and known to all agents.

Each agent $i$ can obtain its accurate state (including position $p_{i,k}$ and velocity $v_{i,k}$) and the control input $u_{i,k}$ at every time step $k \in \mathbb{N}_0$.
From the side of agent $i$, we define
\begin{align}\label{eqn:eqn:AISs of its own}
\mathcal{I}_{i,k}^i : =\left\{ p_{i,0:k}, v_{i,0:k}, u_{i,0:k} \right\}
\end{align}
as the available information set of agent $i$ at time step $k$, where $p_{i,0:k}$, $v_{i,0:k}$, and $u_{i,0:k}$ represent the positions, velocities, and control inputs of agent $i$ up to $k$, respectively.

\begin{figure}[ht]
\centering
\includegraphics [width=0.85\columnwidth]{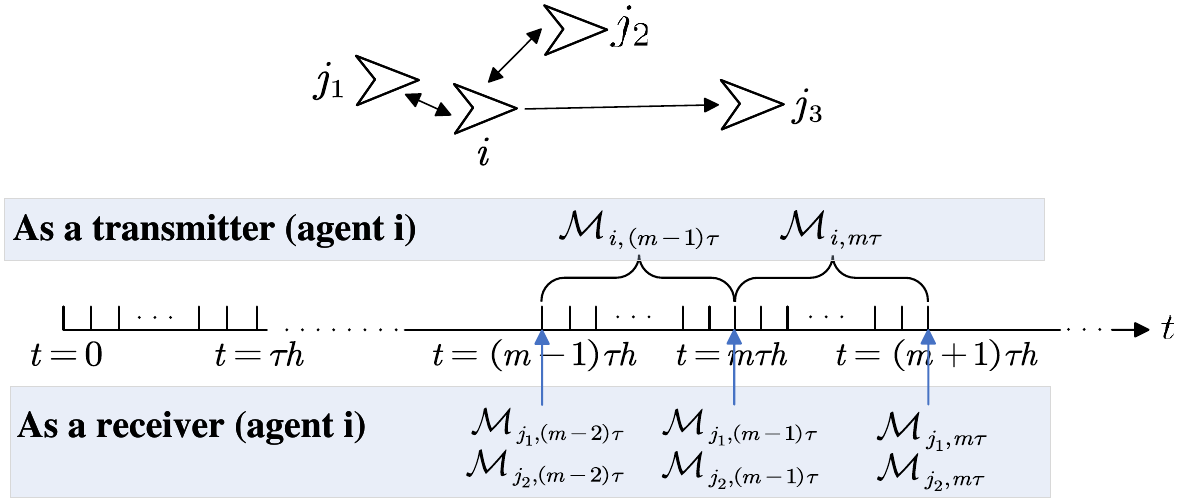}
\caption{
    Illustration of transmitters and receivers.
    Consider agent $i$ with the transmission time $T = \tau h$ can successfully receive messages transmitted by agents $j_1$ and $j_2$.
    At time step $m\tau$, agent $i$ broadcasts message $\mathcal{M}_{i,m\tau}$ and receives messages $\mathcal{M}_{j_1,(m-1)\tau}$ and $\mathcal{M}_{j_2,(m-1)\tau}$ from agents~$j_1$ and $j_2$.
}
\label{fig:TransmitAndReceive}
\end{figure}

As a transmitter, each agent $i$ periodically broadcasts messages with the transmission time $T = \tau h$ ($\tau \in \mathbb{Z}_+$), which can also be called latency~\cite{What_Will_5G_Be}, and the transmit power $P_{i,k}^{\mathrm{tx}}$.
All agents continuously broadcast the messages such that the broadcast period equals the transmission time (see \figref{fig:TransmitAndReceive}).

For each agent $i$, it not only broadcasts messages to its out-neighbors but also receives messages from its in-neighbors, which can be defined as follows.

\begin{definition}[Set of Neighbors]\label{def:Neighbor Sets}
Let $\mathcal{N}_{i}^{\mathrm{in}}(m)$ and $\mathcal{N}_{i}^{\mathrm{out}}(m)$ be the sets of in-neighbors and out-neighbors for agent $i$ at time step $k=m\tau$, where $m \in \mathbb{Z}_+$.
For agents $i,j \in \mathcal{V}~(i\neq j)$ at $k=m\tau$, if:\footnote{For initial time~$k=0$, the sets of in-neighbors and out-neighbors for each agent~$i$, i.e.,~$\mathcal{N}_{i}^{\mathrm{in}}(0)$ and $\mathcal{N}_{i}^{\mathrm{out}}(0)$, are predetermined.}
\begin{itemize}
    \item[(i)] agent $i$ receives messages from agent $j$, then agent $j$ is an in-neighbor of agent $i$ and satisfies $j \in \mathcal{N}_{i}^{\mathrm{in}}(m)$;
    \item[(ii)] agent $j$ receives messages from agent $i$, then agent $j$ is an out-neighbor of $i$ and satisfies $j \in \mathcal{N}_{i}^{\mathrm{out}}(m)$.
\end{itemize}
\end{definition}

With \defref{def:Neighbor Sets}, for a transmitter agent $i$, we consider each message $\mathcal{M}_{i,k}$ includes the position, the velocity, and the function of states from its in-neighbors as follows
\begin{equation}\label{eqn:Message}
\mathcal{M}_{i,k} := \{p_{i,k}, v_{i,k}, f_i(p_{\mathcal{N}_i^{\mathrm{in}}(m), k}, v_{\mathcal{N}_i^{\mathrm{in}}(m), k})\},~k=m \tau,
\end{equation}
where $p_{\mathcal{N}_i^{\mathrm{in}}(m), k}$ is a collection of the known positions for any agent $j \in \mathcal{N}_i^{\mathrm{in}}(m)$ at time $k$, the same as $v_{\mathcal{N}_i^{\mathrm{in}}(m), k}$.
Especially, for the initial time~$k=0$,~$\mathcal{M}_{i,0}:=\bigcup _{l \in \mathcal{V}} \{ p_{l,0}, v_{l,0}\}$.
Moreover, the packet length $M$ is constant and independent of indices $i$ and $k$.

The data rate $\mu$ is the ratio of packet length $M$ to transmission time $T$, i.e.,
\begin{equation}\label{eqn:Data Rate}
\mu=\frac{M}{T}.
\end{equation}
In this work, we assume all agents have the same data rate, and the data rate is fixed during each transmission.

As a receiver, each agent $i$ receives messages from its in-neighbors.
The message $\mathcal{M}_{j,(m-1)\tau}$ transmitted by agent ~$j \in \mathcal{N}_{i}^{\mathrm{in}}(m)$ can only be decoded by agent $i$ at time step $k=m\tau$, i.e., after it has been fully received~\cite{9031321}.
Therefore, from the side of agent $i$, the available information set of its in-neighbor agent $j$ up to time step $k~(k \geq \tau)$ is
\begin{equation}\label{eqn:AISs of in-neighbors}
\mathcal{I}_{\mathcal{N}_{i}^{\mathrm{in}}(\lfloor {k} / {\tau} \rfloor),k}^i : =\bigcup_{l = 1}^{\lfloor {k} / {\tau} \rfloor} \mathcal{I}^i_{j,l\tau} := \bigcup_{l = 1}^{\lfloor {k} / {\tau} \rfloor} \bigcup_{j \in \mathcal{N}_{i}^{\mathrm{in}}(l)}\mathcal{M}_{j,(l-1)\tau}.
\end{equation}
In addition, for~$0 \leq k < \tau$,~$\mathcal{I}_{\mathcal{N}_{i}^{\mathrm{in}}(0),k}^i=\mathcal{M}_{i,0}$.

In communication theory, a message transmitted through a communication channel can be correctly decoded if the data rate $\mu$ is less than the channel capacity $C$, i.e., $\mu < C$.
In the MASs, the channel capacity from agent $i$ (as a transmitter) to agent $j$ (as a receiver) at time step $k$ is~\cite{A_mathematical_theory_of_communication}
\begin{align}\label{eqn:Channel Capacity}
C_{j,i,k}=B_w\log_{2}(1+\varUpsilon_{j,i,k}),
\end{align}
where $B_w$ is the communication bandwidth which is assumed to be a constant, and $\varUpsilon_{j,i,k}$ is the Signal-to-Noise Ratio (SNR) for agent $j$ as follows\footnote{Each agent broadcasts messages using a unique communication band, which indicates the transmitted signals do not interfere each other.} 
\begin{align}\label{eqn:SNR}
\varUpsilon_{j,i,k}=\frac{P_{i,k}^{\mathrm{tx}}g(\|p_{i,k}-p_{j,k}\|)}{W_{j,k}},
\end{align}
where $W_{j,k}$ is the power of noise at agent $j$ and $g(\|p_{i,k}-p_{j,k}\|)$ is the power gain determined by the path loss\footnote{The path loss function in~\cite{Rappaport1996} is defined as the difference (in a logarithmic form) between the effective transmit power and the received power, which is equivalent to our power gain.}
\begin{align}\label{eqn:path loss function}
g(\|p_{i,k}-p_{j,k}\|) = g_{d_0} \frac{d_0^{\psi}}{\|p_{i,k}-p_{j,k}\|^{\psi}},
\end{align}
where $\|p_{i,k}-p_{j,k}\|$ returns the distance between agents $i$ and $j$;
$d_0$ is the reference distance;
$g_{d_0}$ is the power gain at a distance $d_0$ from agent $i$, which depends on antenna gain and wavelength, etc.~\cite{Rappaport1996};
$\psi$ is the path loss exponent.
With~$\mu < C$ and~\eqref{eqn:Channel Capacity}, the condition of SNR for successful transmission is
\begin{equation}\label{eqn:SNR Condition}
\varUpsilon_{j,i,k} > 2^{\frac{\mu}{B_w}}-1.
\end{equation} 

Then, we give the following definition to characterize the communication region of each transmitter agent with a known position.

\begin{definition}[Communication Region]\label{def:Communication Range of Point}
At time step $k \in \mathbb{N}_0$, the communication region of a transmitter agent $i$ located at $p_{i,k}$ is the set of all possible positions of receiver agent $j$ such that the SNR condition for successful transmission can be satisfied, i.e.,
\begin{align}\label{eqn:Communication Range of Point}
\Omega_{p_{i,k}}:=\left\{p_{j,k}\colon \mu < C_{j,i,k} \right\}.
\end{align}
\end{definition}

\begin{figure}[ht]
\centering
\includegraphics [width=0.82\columnwidth]{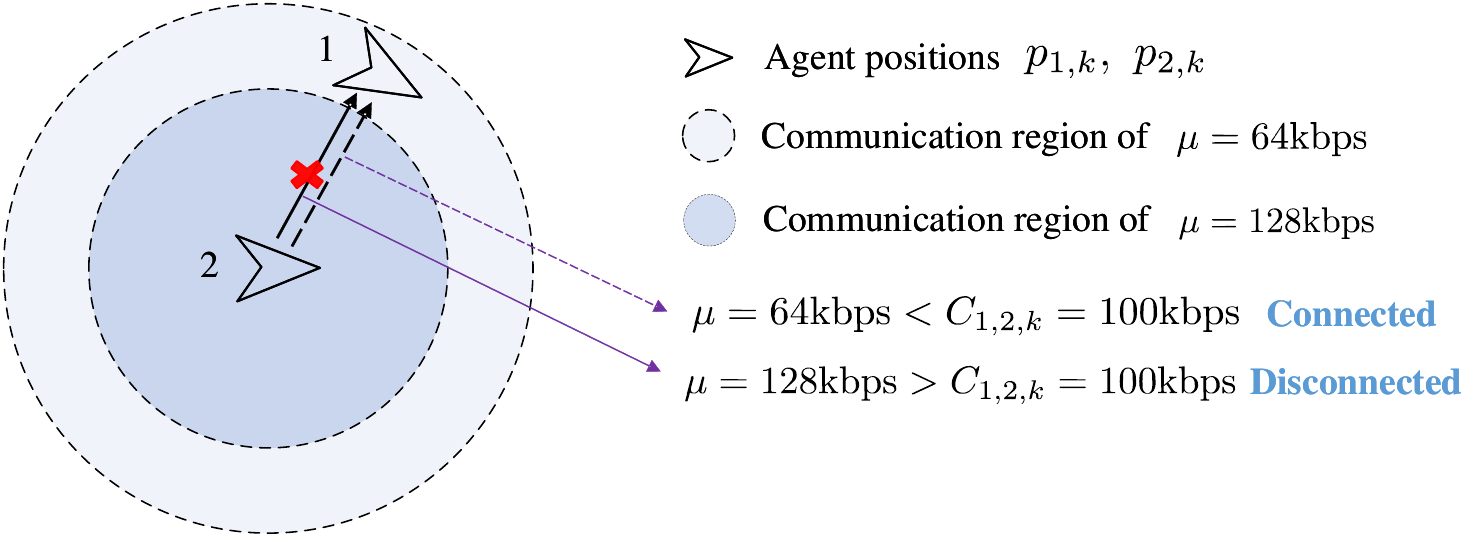}
\caption{
    Illustration of the communication region.
    The channel capacity from agent $2$ to agent $1$ at time $k$ is $C_{1,2,k}=100\mathrm{kbps}$.
    If agent $2$ uses the data rate $\mu=64\mathrm{kbps} < C_{1,2,k}=100\mathrm{kbps}$, agent $1$ can successfully receive (decode) the transmitted messages;
    the corresponding communication region is the (lighter blue) circle with the dashed boundary.
    If agent $2$ uses the data rate $\mu=128\mathrm{kbps} > C_{1,2,k}=100\mathrm{kbps}$, the communication link from agent $2$ to agent $1$ breaks;
    the corresponding communication region is the (darker blue) circle with the dashed boundary.
    It is clear that agent $1$ is within the communication region of agent $2$ if $\mu=64\mathrm{kbps}$, but it falls outside the region if $\mu=128\mathrm{kbps}$.}
\label{fig:CommunicationInteraction}
\end{figure}

With \eqref{eqn:Channel Capacity}-\eqref{eqn:Communication Range of Point}, we can obtain 
\begin{align}\label{eqn:Communication Range Set}
\!\!\!\!\Omega_{p_{i,k}}\!\!=\!\!\left\{p_{j,k}\colon \!\! \|p_{i,k}-p_{j,k}\| \!\! < \!\! d_0\left( \frac{g_{d_0}P_{i,k}^{\mathrm{tx}}}{(2^{\frac{\mu}{B_w}}-1)W_{j,k}} \right)^{\frac{1}{\psi}} \right\}.
\end{align}
The communication region of each transmitter agent with deterministic position is an $n$-dimensional open ball $\Omega_{p_{i,k}}={\mathcal{B}}(c_{i,k},R_{i,k})$, where the center $c_{i,k}$ and the radius $R_{i,k}$ are
\begin{align}\label{eqn:Communication Center}
c_{i,k}&=p_{i,k}, \\ \label{eqn:Communication Radius}
R_{i,k}&=d_0\left( \frac{g_{d_0}P_{i,k}^{\mathrm{tx}}}{(2^{\frac{\mu}{B_w}}-1)W_{j,k}} \right)^{1/\psi}.
\end{align}
Note that the communication region varies with the data rate (determined by the message and the transmission time) and the transmit power, which is usually overlooked in the control community;
an illustrative example is given in \figref{fig:CommunicationInteraction}.

\begin{figure}[ht]
\centering
\includegraphics [width=0.77\columnwidth]{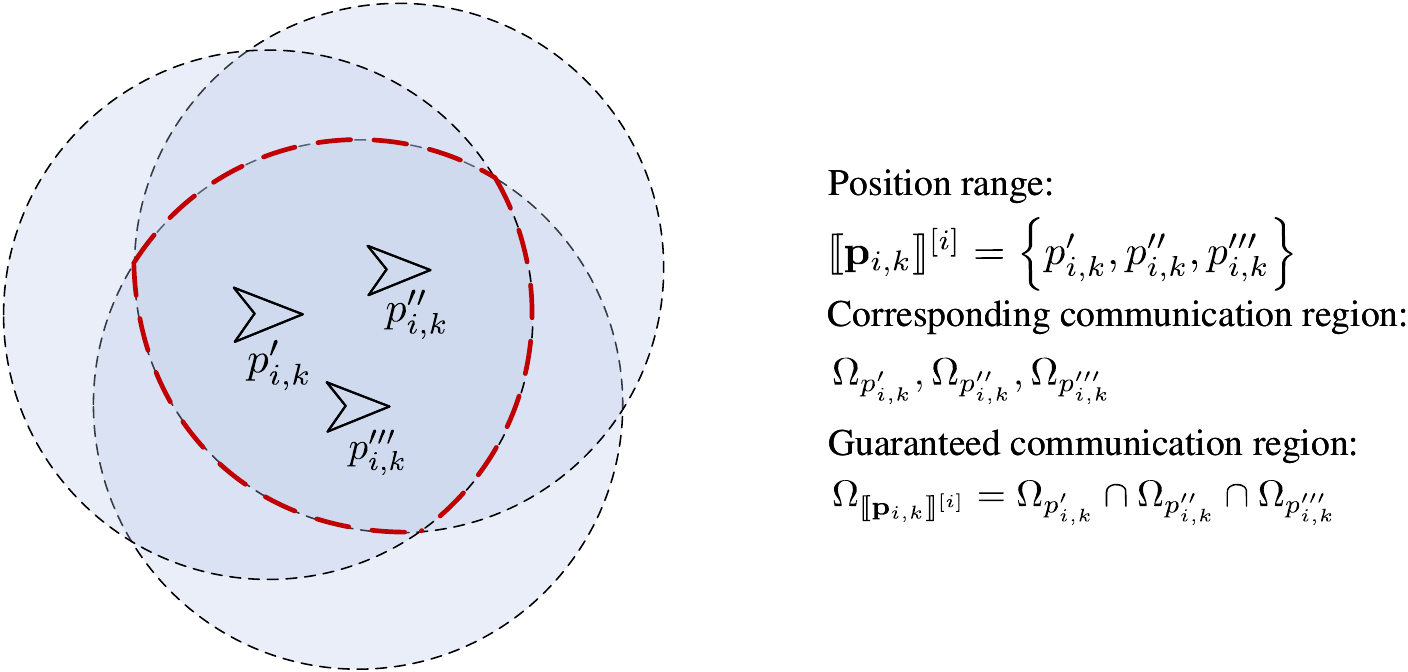}
\caption{
    Illustration of the guaranteed communication region of agent $i$ with position range $\llbracket \mathbf{p}_{i,k}\rrbracket^{[i]}=\left\{ p'_{i,k},p''_{i,k},p'''_{i,k}  \right\}$.
    The communication regions of agent $i$ located at $p'_{i,k}, p''_{i,k}, p'''_{i,k}$ are $\Omega_{p'_{i,k}}, \Omega_{p''_{i,k}}, \Omega_{p'''_{i,k}}$, respectively.
    The guaranteed communication region for agent~$i$ with position range $\llbracket \mathbf{p}_{i,k}\rrbracket^{[i]}$ is the intersection of these (blue) circles, i,e., $\Omega _{\llbracket \mathbf{p}_{i,k} \rrbracket^{[i]}}=\Omega _{p'_{i,k}}\cap \Omega _{p''_{i,k}}\cap \Omega _{p'''_{i,k}}$ (the region enclosed by red dashed lines).
}
\label{fig:IllustrativeOfCommunicationRadius}
\end{figure}

\begin{remark}
In recent studies, the communication links between agents were still assumed to be location independent~\cite{Ma2023,Liu2024,Lin2024} or only location dependent~\cite{Circle_Formation_Control_of_Mobile_Agents_With_Limited_Interaction_Range,Santilli2022}.
All these studies simplify the communication model and ignore the essential constraints of physical layer on communication links.
In contrast, the communication region described in \defref{def:Communication Range of Point} depends on locations, bandwidth, transmit power, and data rate, which reflects the relationship between the controlled movement and communication constraints.
\end{remark}

Since the message can only be decoded after it has been fully received, it is necessary to ensure that the receiver agent is always within the communication region of the transmitter agent during message transmission.
For instance, the conditions that the message $\mathcal{M}_{i,m\tau}$ transmitted by agent $i$ at time~$k=m\tau$ can be successfully received by agent $j$ are
\begin{equation}\label{eqn:conditions of successfully received}
p_{j,m\tau+l} \in \Omega_{p_{i,m\tau+l}},~m \in \mathbb{N}_0,~ l = 0, \ldots, \tau-1.
\end{equation}
Note that the conditions \eqref{eqn:conditions of successfully received} are supposed to be guaranteed from the beginning time of each message transmission.
However, due to the bounded process noises, it is necessary to consider all possible positions of the transmitter agent during message transmission.
From the side of agent~$i$, the prediction of its own position range can be expressed as follows
\begin{equation}
\llbracket\mathbf{p}_{i,k}\rrbracket^{[i]} : =\llbracket\mathbf{p}_{i,k} | \mathcal{I}_{i,\lfloor k/\tau \rfloor\tau}^{i}, \mathcal{I}_{\mathcal{N}_{i}^{\mathrm{in}}(\lfloor {k} / {\tau} \rfloor),k}^i \rrbracket,
\end{equation}
where $\mathcal{I}_{i,\lfloor k/\tau \rfloor\tau}^{i}$ and $\mathcal{I}_{\mathcal{N}_{i}^{\mathrm{in}}(\lfloor {k} / {\tau} \rfloor),k}^i$ are defined by \eqref{eqn:eqn:AISs of its own} and \eqref{eqn:AISs of in-neighbors}.

Based on \defref{def:Communication Range of Point}, we give the following definition to characterize the guaranteed communication region of the transmitter agent with position uncertainty.

\begin{definition}[Guaranteed Communication Region]\label{def:Communication Range With Uncertain}
 At time step $k \in \mathbb{N}_0$, the guaranteed communication region of a transmitter agent $i$ with position range $\llbracket\mathbf{p}_{i,k}\rrbracket^{[i]}$ is the intersection of the communication regions for agent $i$ located at all possible positions satisfying $p_{i,k} \in \llbracket \mathbf{p}_{i,k}\rrbracket^{[i]}$, i.e.,
\begin{equation}
\Omega_{\llbracket\mathbf{p}_{i,k}\rrbracket^{[i]}}=\bigcap\limits_{p_{i,k} \in \llbracket \mathbf{p}_{i,k}\rrbracket^{[i]}} \Omega_{p_{i,k}}.
\end{equation}
\end{definition}

An illustration is given in \figref{fig:IllustrativeOfCommunicationRadius}, where the position range consists of three separated points.
In general, the position range $\llbracket\mathbf{p}_{i,k}\rrbracket^{[i]}$ is an uncountable set of all possible positions.

The guaranteed communication region determines the communication topology.
A directed graph $\mathcal{G}(m)=(\mathcal{V},\mathcal{E}(m),\mathcal{A}(m))$, where~$m=\lfloor k/\tau \rfloor \in \mathbb{N}_0$, can be utilized to model the time-varying communication interactions among agents at time step $k$, where node $i \in \mathcal{V} = \{1,\ldots,N\}$ stands for the $i$\textsuperscript{th} agent, and the edge $(i,j) \in \mathcal{E}(m) \subseteq \mathcal{V} \times \mathcal{V}$ represents for the communication link from agent $i$ to agent $j$.
The adjacency matrix $\mathcal{A}(m)$ satisfies: $a_{ji} = 1$ if and only if $(i,j) \in \mathcal{E}(m)$; otherwise, $a_{ji} = 0$.
Then, we define the edge set based on the guaranteed communication region as follows.

\begin{definition}[Edge Set]\label{def:Edge Set}
For agents $i, j \in \mathcal{V}$ ($i \neq j$), an edge $(i,j) \in \mathcal{E}(m+1)~(m \in \mathbb{N}_0)$ if
\begin{equation}\label{eqn:conditions of successfully received-with uncertainty}
p_{j,m\tau+l} \in  \Omega_{\llbracket \mathbf{p}_{i,m\tau+l} \rrbracket^{[i]}},~ l = 0, \ldots, \tau-1.
\end{equation}
In addition,~$\mathcal{E}(0)$ depends on the in-neighbors set $\mathcal{N}_{i}^{\mathrm{in}}(0)$ and the out-neighbors set $\mathcal{N}_{i}^{\mathrm{out}}(0)$ for~$i \in \mathcal{V}$.
\end{definition}

Then, we can derive the sets of in-neighbors and out-neighbors for agent $i$ as follows.
\begin{lemma}[Set of Neighbors]\label{lem:Neighbor Sets}
The sets of in-neighbors and out-neighbors of agent $i$ with position range~$\llbracket \mathbf{p}_{i,k}\rrbracket^{[i]}$ are
\begin{equation}\label{eqn:set of in-neighbor}
	\mathcal{N}_i^{\mathrm{in}}(m) = \{j_1 \colon (j_1,i) \in \mathcal{E}(m),~j_1 \in \mathcal{V}\},
\end{equation}
\begin{equation}\label{eqn:set of out-neighbor}
	\mathcal{N}_i^{\mathrm{out}}(m) = \{j_2 \colon (i,j_2) \in \mathcal{E}(m),~j_2 \in \mathcal{V}\},
\end{equation}
respectively, where $m \in \mathbb{Z}_+$.
\end{lemma}

\begin{IEEEproof}
According to~\defref{def:Neighbor Sets} and~\defref{def:Edge Set}, the sets of neighbors can be easily obtained; see~\eqref{eqn:set of in-neighbor} and~\eqref{eqn:set of out-neighbor}.\hfill
\end{IEEEproof}

\subsection{Problem Description}\label{sec:Problem Description}

In this work, we focus on the formation control problem with the communication model in \secref{sec:System Model}, which includes: (i) analyzing the fundamental limits of data rate under a given control accuracy level (see \probref{prob:First}), and (ii) providing an integrated design of control and communication (see \probref{prob:Second}). 
The details are as follows.

The formation pattern is predetermined by formation vector $\Delta=[\Delta_{1}^{T},\Delta_{2}^{T},\ldots,\Delta_{N}^{T}]^{T} \in \mathbb{R}^{nN}$, and $\Delta_{i,j}=\Delta_{i}-\Delta_{j}$ denotes the desired relative position of agent~$i$ w.r.t. agent~$j$.
The MASs \eqref{eqn:Second-Order of Position and Velocity} are said to achieve desired formation pattern if 
\begin{equation}\label{eqn:formation pattern achievement}
\underset{k \to \infty}{\lim } \left(\begin{bmatrix} {p}_{i,k} \\ {v}_{i,k} \end{bmatrix} -\begin{bmatrix} {p}_{j,k} \\ {v}_{j,k} \end{bmatrix}\right) = \begin{bmatrix}\Delta_{i,j} \\ \mathbf{0}_n \end{bmatrix},~\forall i,j \in \mathcal{V}.
\end{equation}

The uncertainties introduced by the bounded process noises and the discontinuous state information of neighbors caused by the finite data rate bring negative effects to the control accuracy.
To describe the formation accuracy, we define $\boldsymbol{\delta}_{i,j,k}^p$ and $\boldsymbol{\delta}_{i,j,k}^v$ as the position and velocity formation errors of agent $i$ w.r.t. agent $j$ at time $k$ with their realizations $\delta_{i,j,k}^p \in \llbracket \boldsymbol{\delta}_{i,j,k}^p \rrbracket \subseteq \mathbb{R}^n$ and $\delta_{i,j,k}^v \in \llbracket \boldsymbol{\delta}_{i,j,k}^v \rrbracket \subseteq \mathbb{R}^n$, respectively:
\begin{align}
\boldsymbol{\delta}_{i,j,k}^p&:=\mathbf{p}_{i,k}-\mathbf{p}_{j,k}-\Delta_{i,j}, \label{eqn:Relative Position}\\
\boldsymbol{\delta}_{i,j,k}^v&:=\mathbf{v}_{i,k}-\mathbf{v}_{j,k}. \label{eqn:Relative Velocity}
\end{align}

Then, the bounded formation of MASs \eqref{eqn:Second-Order of Position and Velocity} can be defined by the boundedness of the position/velocity formation errors.

\begin{definition}[Bounded Formation Stability]\label{def:Bounded formation stability}
The formation of MASs described by~\eqref{eqn:Second-Order of Position and Velocity} is bounded, for $\forall i,j \in \mathcal{V}$, if
\begin{align}
\limsup_{k\to \infty} \|\delta_{i,j,k}^{p}\| & \leq \delta_p,\label{eqn:Formation Error Bound of Position}\\
\limsup_{k\to \infty} \|\delta_{i,j,k}^{v}\| & \leq \delta_v,\label{eqn:Formation Error Bound of Velocity}
\end{align}
where $\delta_p,~\delta_v \in \mathbb{R}$ are the steady-state formation error bounds of position and velocity for MASs, respectively.
Or equivalently, we say the MASs have achieved bounded formation stability.
\end{definition}

In \defref{def:Bounded formation stability}, the formation error bounds reflect the control accuracy of a formation in a steady-state manner.
The control accuracy is not only affected by the inevitable process noises but also depends on the frequency of information exchange among agents.

On the one hand, the data rate cannot be arbitrarily high due to the constraints of the communication physical layer; on the other hand, the data rate cannot be too low since the discontinuous feedback state information caused by latency brings negative effects to the formation control for MASs.

Therefore, it is necessary to analyze the fundamental limits of data rate for MASs with the communication model described in \secref{sec:System Model} and a given control accuracy; see \probref{prob:First} as follows.

\begin{problem}[Fundamental Limits of Data Rate]\label{prob:First}
How to characterize the fundamental limits (i.e., upper and lower bounds) of data rate for MASs to achieve bounded formation with any given control accuracy?
\end{problem}

Note that \probref{prob:First} is considered in a steady-state manner, which means the channel capacities suffice to support the data rates in desired links when $k$ is sufficiently large.
However, during the transient process, some desired links may break due to insufficient channel capacities.
To ensure successful transmission and save energy\footnote{On the one hand, the transmit power should be large enough to provide a suitable guaranteed communication region, which ensures the connections of desired communication links;
on the other hand, the transmit power cannot be too large, which results in significant power consumption.}, it is necessary to efficiently guarantee the channel capacity to support the data rate in \probref{prob:First} by controlling the transmit power; meanwhile, the control law should be carefully designed to adapt the communication.
Thus, it is significant to jointly consider formation control and power control such that the MASs can achieve bounded formation.

The power control strategy of MASs \eqref{eqn:Second-Order of Position and Velocity} can be formally defined as follows.

\begin{definition}[Power Control Strategy]\label{def:power control}
For a transmitter agent $i \in \mathcal{V}$ with dynamics \eqref{eqn:Second-Order of Position and Velocity}, the power control strategy is a sequence of transmit power $P_{i,k}^{\mathrm{tx}}~(k \in \mathbb{N}_0)$ such that the guaranteed communication region can be adaptively adjusted.
\end{definition}

With \defref{def:power control}, we propose \probref{prob:Second} as follows.

\begin{problem}[Integrated Design of Control and Communication]\label{prob:Second}
How to design the formation control law and the power control strategy jointly for MASs to achieve bounded formation with a given control accuracy?
\end{problem}

Then, we give the theoretical analysis of the fundamental limits for data rate in \secref{sec:Analysis of Fundamental Limit for Data Rate}, and put forward an integrated design of formation control and power control in \secref{sec:Closed-Loop Design of Formation Control and Communication Transmission}.

\section{Analysis of Fundamental Limits of Data Rate}\label{sec:Analysis of Fundamental Limit for Data Rate}

In this section, firstly we analyze the predicted position range to describe the position uncertainty for agents (see \secref{sec:Analysis of Uncertain Position Range}).
Then, we establish the relationship between the guaranteed communication radius, the transmit power, and the transmission time, and derive the property of the guaranteed communication radius (see \secref{sec:Property of Guaranteed Communication Radius}).
Finally, we characterize the fundamental limits of the data rate (see \secref{sec:Fundamental Limit for Data Rate}).

\subsection{Analysis of Position Range}\label{sec:Analysis of Uncertain Position Range}

With \defref{def:Communication Range With Uncertain}, the guaranteed communication region of each transmitter agent varies with the uncertain position range.
Thus, it is necessary to analyze the position range which is characterized in \propref{prop:Estimation of Uncertain Position Range}.

\begin{proposition}[Position Range]\label{prop:Estimation of Uncertain Position Range}
For each agent $i \in \mathcal{V}$ with dynamics \eqref{eqn:Second-Order of Position and Velocity}, the position range $\llbracket\mathbf{p}_{i,k}\rrbracket^{[i]} =\llbracket\mathbf{p}_{i,k}|\mathcal{I}^i_{i,k-\sigma}\rrbracket~(\sigma=k-\lfloor k/\tau \rfloor \tau)$ is an $n$-dimensional closed ball $\mathcal{B}[c_p(i,k),r_p(i,k)]$ with the center $c_p(i,k)$ and the radius $r_p(i,k)$:
\begin{align}
\nonumber
c_p(i,k)&=p_{i,k-\sigma}+\sigma hv_{i,k-\sigma} \\
& +h^2\sum_{l=1}^{\sigma}(\sigma+\frac{1}{2}-l)c_u(i,k-\sigma+l-1), \\
\nonumber
r_p(i,k)&=h^2\sum_{l=1}^{\sigma}(\tau+\frac{1}{2}-l)r_u(i,k-\tau+l-1) \\ \label{eqn:radius of position range}
&+\sigma r_{w_p}(i)+\frac{1}{2}\sigma(\sigma-1)hr_{w_v}(i),
\end{align}
for~$\sigma>0$, where $c_u(i,k-\sigma+l-1)$ and $r_u(i,k-\sigma+l-1)$ are the center and the radius of~$\llbracket\mathbf{u}_{i,k-\sigma+l-1}\rrbracket$.\footnote{The unknown range of control input $\mathbf{u}_{i,k}$ is assumed to be an $n$-dimensional closed ball.
If the actual range of $\mathbf{u}_{i,k}$ is not a ball, we can approximate it with the circumscribed ball.}
Specially, $\llbracket\mathbf{p}_{i,k}\rrbracket^{[i]} =\llbracket\mathbf{p}_{i,k}|\mathcal{I}^ir_{i,k}\rrbracket=p_{i,k}$ for $\sigma=0$.
\end{proposition}

\begin{IEEEproof}
See \apxref{apx:Proof of prop:Estimation of Uncertain Position Range}.
\end{IEEEproof}

\begin{remark}
Since the radii of the ranges for the unknown control inputs and the position/velocity process noises are positive, the radius of position range increases as $\sigma$ grows.
\end{remark}

If the uncertainty of the control inputs can be eliminated (see our design and \rekref{rek:Controller uncertainty eliminate} in \secref{sec:Formation Controller Design}), the radius $r_u(i,k-\tau+l-1)$ in \eqref{eqn:radius of position range} is zero.
Then, the corresponding lower bound of the radius for position range at time step $k$ is
\begin{align}\label{eqn:radius without control uncertainty}
r_p(i,k)=\sigma r_{w_p}(i)+\frac{1}{2}\sigma(\sigma-1)hr_{w_v}(i).
\end{align}

\subsection{Property of Guaranteed Communication Radius}\label{sec:Property of Guaranteed Communication Radius}
\begin{figure}
\centering
\subfigure[] {\includegraphics [width=0.7\columnwidth]{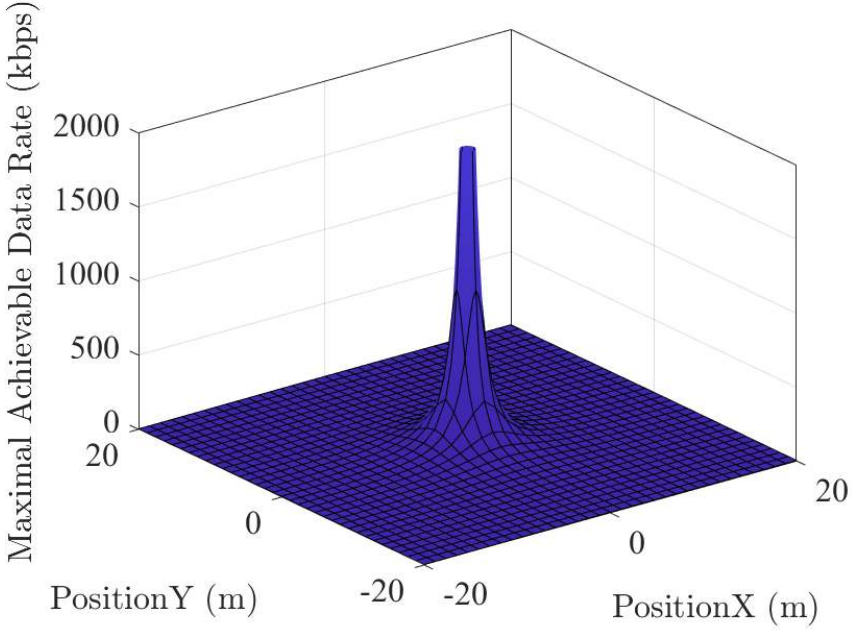}\label{fig:CapacityWithPositionXY}} 
\subfigure[] {\includegraphics [width=0.62\columnwidth]{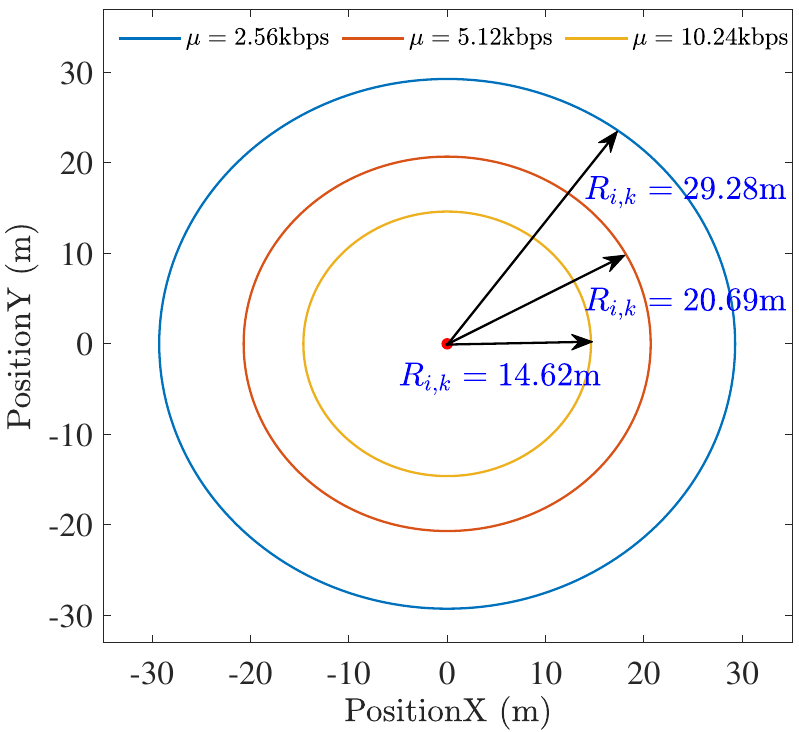}\label{fig:CapacityForDifferentCR}}
\caption{
The relationship between the communication region and the data rate.
Consider two agents $i$ (as a transmitter) and $j$ (as a receiver) moving in X-Y plane.
The sampling period $h=0.05\mathrm{s}$ and the radii of both position and velocity noise ranges for each agent are $0.5\mathrm{m}$.
The communication parameters are: $M=64*8\mathrm{bits},~B_w=10^6\mathrm{Hz},~d_0=1\mathrm{m},~g_{d_0}=1/(16\pi)^2,~P_{i,k}^{\mathrm{tx}}=1\mathrm{w},~\psi=2$.
The noise power of each agent at time instant $k \in \mathbb{N}_0$ is the sum of channel noise and jamming noise, i.e., $W=(N_0+N_{\rm{jam}})B_w$, where the channel and jamming noise power spectrum density are $N_0=10^{-11}\mathrm{w}/\mathrm{Hz}$ and $N_{\rm{jam}}=2.5*10^{-10}\mathrm{w}/\mathrm{Hz}$, respectively.
Take the location of agent $i$ as the origin of the 2-D plane.
(a) The illustration of the maximal achievable data rate w.r.t. the relative position between agents $i$ and $j$.
(b) The communication radii corresponding to different data rates.
}
\label{fig:CapacityCurves}
\end{figure}

With \eqref{eqn:Communication Radius}, the communication radius $R_{i,k}$ of agent $i$ located at $p_{i,k}$ is dependent on data rate.
To satisfy the condition of successful decoding $\mu < C$, we characterize the relationship between the communication region and the data rate in \figref{fig:CapacityCurves}.\footnote{According to the Examples~4.1 and~4.2 in~\cite{Rappaport1996}, we consider: the largest physical linear dimension of the antenna is $0.2\mathrm{m}$; the wavelength is $0.25\mathrm{m}$; the transmitter and receiver antenna gains are $1\mathrm{dBi}$; the system loss factor is $1$. Thus, we select $d_0=1\mathrm{m}$ and $g_{d_0}=1/(16\pi)^2$ in numerical example.}

Based on \defref{def:Communication Range With Uncertain}, we provide the following lemma to obtain the radius of guaranteed communication region for agent~$i$ with position range $\llbracket\mathbf{p}_{i,k}\rrbracket^{[i]}=\mathcal{B}[c_p(i,k),r_p(i,k)]$.

\begin{lemma}[Guaranteed Communication Radius]\label{lem:Guaranteed Communication Radius}
	The guaranteed communication region for agent~$i \in \mathcal{V}$ with position range~$\llbracket\mathbf{p}_{i,k}\rrbracket^{[i]}=\mathcal{B}[c_p(i,k),r_p(i,k)]$ is an $n$-dimensional open ball, i.e., $\Omega_{\llbracket\mathbf{p}_{i,k}\rrbracket^{[i]}}={\mathcal{B}}(c_p(i,k),R_{\llbracket\mathbf{p}_{i,k}\rrbracket^{[i]}})$, where the guaranteed communication radius (GCR) is
	\begin{align}\label{eqn:Communication radius with position range}
		R_{\llbracket\mathbf{p}_{i,k}\rrbracket^{[i]}}=R_{i,k}-r_p(i,k).
	\end{align}
\end{lemma}

\begin{IEEEproof}
	See \apxref{apx:Proof of lem:Guaranteed Communication Radius}.
\end{IEEEproof}

Recall that the conditions of successful decoding, given in~\eqref{eqn:conditions of successfully received-with uncertainty}, need to be satisfied from the beginning time instant of each message transmission, i.e.,~$k=m\tau~(m \in \mathbb{N}_0)$.
Each agent~$i$ is required to predict its position ranges for~$k \in (m\tau,(m+1)\tau)$ for determining the GCR given in~\eqref{eqn:Communication radius with position range}.
According to~\eqref{eqn:radius without control uncertainty}, the radius of position range increases as~$\sigma$ grows when the uncertainties of the control inputs can be eliminated.
Therefore, we consider the worst-case with the maximum position uncertainty, i.e., $\sigma=\tau-1$.
Then, we give the property of the GCR for each transmitter agent at $k'=m\tau-1$ ($m \in \mathbb{Z}_+$) as follows.

\begin{proposition}[Property of Guaranteed Communication Radius]\label{prop:Concave Property}
For a transmitter agent $i \in \mathcal{V}$ with dynamics \eqref{eqn:Second-Order of Position and Velocity} and transmit power $P_{i,k}^{\mathrm{tx}}$, given $r_u(i,k)=0$, if the communication bandwidth $B_w$ satisfies
\begin{align}\label{eqn:D_2<0}
    2\tau hB_w \psi \big ( 1-2^{\frac{M}{\tau hB_w}} \big )+M\ln2\big(\psi+2^{\frac{M}{\tau hB_w}}\big)<0,
\end{align}
the GCR at time step $k'=m\tau-1$ ($m \in \mathbb{Z}_+$) has the following properties:

\begin{itemize}
    \item [(i)] $R_{\llbracket\mathbf{p}_{i,k'}\rrbracket^{[i]}}$ is concave w.r.t. $\tau$;

    \item [(ii)] $R_{\llbracket\mathbf{p}_{i,k'}\rrbracket^{[i]}}$ increases monotonically for $\tau \in (0,\mathring{\tau}_{i,k'}]$, and decreases monotonically for $\tau \in (\mathring{\tau}_{i,k'},\infty)$, where $\mathring{\tau}_{i,k'}$ is the solution to $dR_{\llbracket\mathbf{p}_{i,k'}\rrbracket^{[i]}}/d\tau=0$.
\end{itemize}
\end{proposition}

\begin{IEEEproof}
See \apxref{apx:Proof of prop:Concave Property}.
\end{IEEEproof}

\begin{remark}\label{rek:bandwidth limit}
For~\eqref{eqn:D_2<0}, we can derive an important limit:
\begin{equation*}
\begin{split}
&\lim_{B_w \to \infty} \Big[2\tau h B_w \psi \big ( 1-2^{\frac{M}{\tau hB_w}} \big )+M\ln2 \big(\psi+2^{\frac{M}{\tau hB_w}}\big)\Big] \\
=& -M\ln2(\psi-1).
\end{split}
\end{equation*}
It indicates that when $\psi>1$\footnote{The path loss exponent $\psi$ depends on the specific propagation environment and usually satisfies $\psi \geq 2$, where $\psi=2$ if and only if the radio propagates in free space~\cite{Rappaport1996}. This means $\psi>1$ is generally observed.}, $\exists \underline{B} >0$ such that the properties~(i) and~(ii) in \propref{prop:Concave Property} hold.
\end{remark}

\begin{figure}[ht]
\centering
\subfigure[] {\includegraphics [width=0.7\columnwidth]{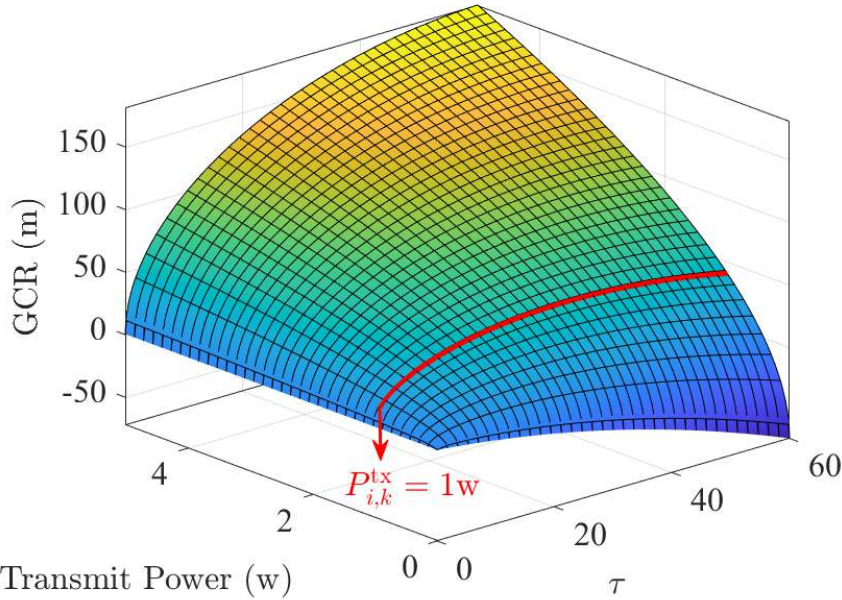}\label{fig:CommunicationRadius}} 
\subfigure[] {\includegraphics [width=0.6\columnwidth]{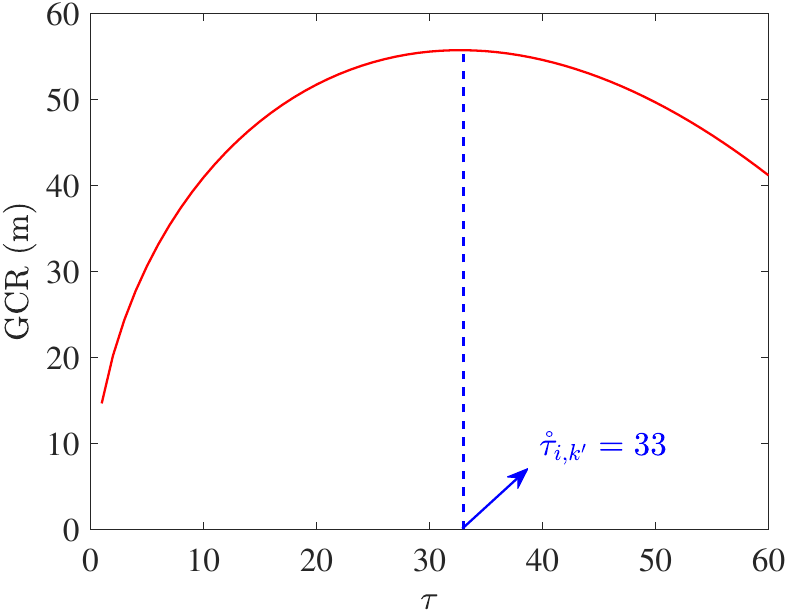}\label{fig:CommunicationRadiusCurveWithFixedPower}}
\caption{
    Illustration of \propref{prop:Concave Property} for agent $i$.
    The parameters are the same as those in \figref{fig:CapacityCurves}.
    (a) The relationship between the GCR, the transmit power, and $\tau$.
    (b) Curve of the GCR w.r.t $\tau$, where the transmit power is $P_{i,k}^{\mathrm{tx}}=1\mathrm{w}$.
}
\label{fig:CommunicationRadiusCurve}
\end{figure}

As shown in \figref{fig:CommunicationRadiusCurve}, we characterize the relationship between the GCR, the transmit power, and the transmission time.
With $\mu=M/(\tau h)$ and \propref{prop:Concave Property}, we can conclude that decreasing the data rate (i) increases the GCR for $\tau \in [1,\mathring{\tau}_{i,k'}]$; (ii) does not improve the GCR for $\tau \in (\mathring{\tau}_{i,k'},\infty)$ due to the significant increased radius of position range.
Therefore,~$\tau$ is better not to exceed~$\mathring{\tau}_{i,k'}$ since sacrificing the data rate will reduce the GCR instead.

Note that the concave property of the GCR in \propref{prop:Concave Property} holds when $B_w > \underline{B}$.
The communication radii (without location uncertainties) corresponding to different data rates and communication bandwidths are given in \tabref{tab:CRAndBandwidth}.
From \tabref{tab:CRAndBandwidth}, we can see that the communication radius decreases with the bandwidth.
The properties in \propref{prop:Concave Property} still hold even for $B_w = 10^3\mathrm{Hz}$.
Since for most wireless communication systems, the bandwidth is typically greater than $10^5\mathrm{Hz}$~\cite{Narrowband_Internet_of_Things_Evolutions_Technologies_and_Open_Issues}, it is reasonable to consider~\eqref{eqn:D_2<0} is satisfied in general.

According to the above analysis, fundamental limits exist in determining the data rate.
On the one hand, based on \propref{prop:Estimation of Uncertain Position Range}, a low data rate (high latency) increases the position/velocity uncertainties of agents and makes it difficult to meet the requirement of steady-state errors; on the other hand, the data rate cannot be arbitrarily high since the limited channel capacities among agents are not enough to support it.

Thus, it is necessary to analyze the fundamental limits of the data rate to provide a guideline for transmission.

\begin{table}
\begin{center}
\caption{The communication radius $R_{i,k}$ for agent $i$ with $P_{i,k}^{\mathrm{tx}}=1\mathrm{w}$}
\label{tab:CRAndBandwidth}
\begin{tabular}{*{4}{c}}
\toprule

& $\mu=5.12\mathrm{kbps}$ & $\mu=10.24\mathrm{kbps}$ \\
\midrule
$B_w=10^3\mathrm{Hz}$ & $6.71\mathrm{m}$ & $1.12\mathrm{m}$ \\
$B_w=10^4\mathrm{Hz}$ & $18.90\mathrm{m}$ & $12.14\mathrm{m}$ \\
$B_w=10^5\mathrm{Hz}$ & $20.53\mathrm{m}$ & $14.39\mathrm{m}$ \\
$B_w=10^6\mathrm{Hz}$ & $20.69\mathrm{m}$ & $14.62\mathrm{m}$ \\
\bottomrule
\end{tabular}
\end{center}
\end{table}

\subsection{Fundamental Limits of Data Rate}\label{sec:Fundamental Limit for Data Rate}

In this subsection, we provide a theorem and a corollary to characterize the fundamental limits of the data rate for achieving the desired bounded formation, i.e.,~$\forall i,j \in \mathcal{V}$ ,~$\limsup_{k\to \infty}\|\delta_{i,j,k}^p\| \leq \delta_p$,~$\limsup_{k\to \infty}\|\delta_{i,j,k}^v\| \leq \delta_v$.\footnote{In the following analysis of this subsection, we consider each agent $i$ receives messages transmitted from agent $j \in \mathcal{V}\backslash\{i\}$ and estimates their position ranges, which implies a more fundamental analysis.}


\begin{theorem}[Fundamental Limits of Data Rate]\label{thm:feasible range of data rate}
For a transmitter agent $i \in \mathcal{V}$ with dynamics \eqref{eqn:Second-Order of Position and Velocity}, given the transmit power $P_{i,k}^{\mathrm{tx}}$ and the formation error bound of position $\delta_p$, the feasible set of data rate at time $k$ is
\begin{align}
\mathcal{F}_{i,k}=\left\{ \frac{M}{\tau h}\colon \Psi_{i,k}^1(\tau) \geq 0,~\Psi_{i,k}^2(\tau) \geq 0,~\tau \in \mathbb{Z}_+ \right\},
\end{align}
where 
\begin{align}
\Psi_{i,k}^1(\tau)& =R_{\llbracket\mathbf{p}_{i,k}\rrbracket^{[i]}}-\max_{j \in \mathcal{N}^{\mathrm{out}}_{i}(\lfloor k/\tau \rfloor)}(\|\Delta_{i,j}\|+r_p(j,k)), \\
\Psi_{i,k}^2(\tau)& =\delta_p -r_p(i,k) -\max_{j \in \mathcal{V}\backslash\{i\}}{r_p(j,k)}.
\end{align}
\end{theorem}


\begin{IEEEproof}
On the one hand, the data rate cannot be arbitrarily low to ensure that the given control accuracy can be achieved.
Denote the position ranges of agents $i$ and $j$ at time $k$ as $\llbracket\mathbf{p}_{i,k}\rrbracket^{[i]}=\mathcal{B}[c_p(i,k),r_p(i,k)]$ and $\llbracket\mathbf{p}_{j,k}\rrbracket^{[i]}=\mathcal{B}[c_p(j,k),r_p(j,k)]$, respectively, where $\llbracket\mathbf{p}_{j,k}\rrbracket^{[i]}$ is the estimate position range of agent $j$ from the side of agent $i$.
According to the Minkowski sum of balls defined in \apxref{apx:Proof of prop:Estimation of Uncertain Position Range}, we have
\begin{equation}
\llbracket \boldsymbol{\delta}_{i,j,k}^p \rrbracket^{[i]} =\llbracket \mathbf{p}_{i,k} \rrbracket^{[i]} \oplus \llbracket -\mathbf{p}_{j,k} \rrbracket^{[i]} \oplus \{ -\Delta_{i,j}\},
\end{equation}
which can be denoted by an $n$-dimensional closed ball $\mathcal{B}[c_p(i,k)-c_p(j,k)-\Delta_{i,j},r_p(i,k)+r_p(j,k)]$.
To satisfy the condition: $\limsup_{k\to \infty}\|\delta_{i,j,k}^p\| \leq \delta_p$, we have
\begin{align}\label{eqn:ri+rj}
r_p(i,k)+r_p(j,k) \leq \delta_p,~j \in \mathcal{V}\backslash\{i\}.
\end{align}

On the other hand, the data rate cannot be arbitrarily high, which indicates a large channel capacity to support it.
However, the distance between a pair of communication agents is constrained by the desired formation pattern and the uncertain position ranges of agents cannot overlap.
From the perspective of the transmitter agent $i$, the GCR should satisfy
\begin{align}\label{eqn:R-2ri-2rj}
R_{\llbracket\mathbf{p}_{i,k}\rrbracket^{[i]}} \geq \| \Delta_{i,j}\|+r_p(j,k),~j \in \mathcal{N}_i^{\mathrm{out}}(\lfloor k/\tau \rfloor).
\end{align}
\end{IEEEproof}

\begin{remark}
	With~\thmref{thm:feasible range of data rate}, as the radii of position ranges for agents increase, we obtain: (i) the minimum solution of~$\tau$ for~$\Psi_{i,k}^1(\tau) \geq 0$ increases, and (ii) the maximum solution of~$\tau$ for~$\Psi_{i,k}^2(\tau) \geq 0$ decreases, which directly affects the fundamental limits of the data rate.
\end{remark}

If the uncertainties of the control inputs can be eliminated, we give a corollary to characterize the feasible set of data rate for each agent~$i$ at time step $k'=m\tau-1$.

\begin{corollary}\label{cor:data rate without controller uncertainty}
For a transmitter agent $i \in \mathcal{V}$ with dynamics \eqref{eqn:Second-Order of Position and Velocity}, transmit power $P_{i,k'}^{\mathrm{tx}}$, and position formation error bound $\delta_p$, given $r_u(i,k)=0$ and $r_u(j,k)=0$ ($j \in \mathcal{V}\backslash\{i\},~k \in \mathbb{N}_0$), if $\psi>1$ and $B_w>\underline{B}$, the feasible set of data rate for time~$k'=m\tau-1$ is
\begin{equation}\!\!\!\!\!\!
\mathcal{F}_{i,k'}=\left\{\frac{M}{\tau h} \colon S^{\min}(\Psi_{i,k'}^1) \leq \tau \leq S^{\max}(\Psi_{i,k'}^2),~\tau \in \mathbb{Z}_+ \right\},
\end{equation}
where
\begin{align}
\nonumber 
\Psi_{i,k'}^1&=d_0\left( \frac{g_{d_0}P_{i,k'}^{\mathrm{tx}}}{(2^{\frac{M}{B_w\tau h}}-1)W_{j,k'}} \right)^{1/\psi}-\max_{j \in \mathcal{N}^{\mathrm{out}}_{i}(m-1)}\|\Delta_{i,j}\| \\
\nonumber 
&-\max_{j \in \mathcal{N}^{\mathrm{out}}_{i}(m-1)}\left[ (2\tau-1)r_{w_p}(j)+(\tau-1)hr_{w_v}(j))
\right] \\ \label{eqn:explicit form-1}
&-[ (\tau-1)(r_{w_p}(i)+\frac{(\tau-2)}{2}hr_{w_v}(i))], \\
\nonumber
\Psi_{i,k'}^2&=\delta_p-\max_{j \in \mathcal{V}\backslash\{i\}}\left[ (2\tau-1)(r_{w_p}(j)+(\tau-1)hr_{w_v}(j))
\right] \\ \label{eqn:explicit form-2}
&-[ (\tau-1)(r_{w_p}(i)+\frac{(\tau-2)}{2}hr_{w_v}(i))].
\end{align}
\end{corollary}

\begin{IEEEproof}
See \apxref{apx:Proof of cor:data rate without controller uncertainty}.
\end{IEEEproof}

\begin{figure}[ht]
\centering
\includegraphics [width=0.6\columnwidth]{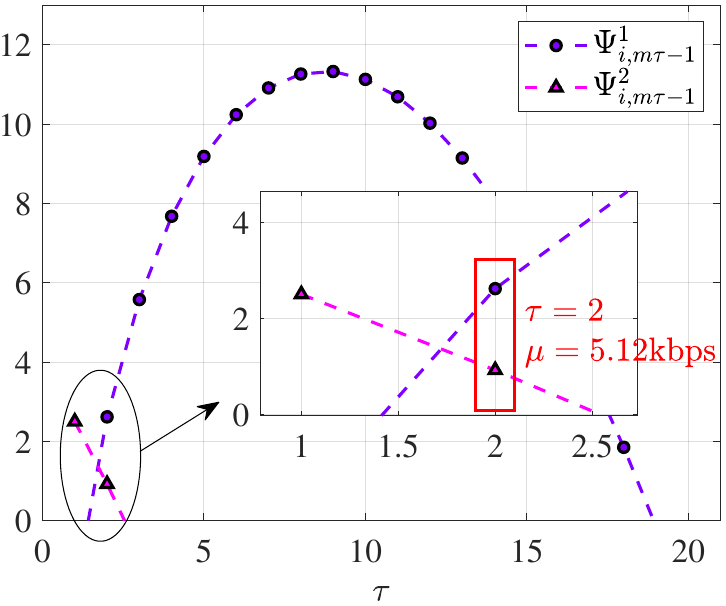}
\caption{
Illustration of~\corref{cor:data rate without controller uncertainty}.
Let $\delta_p=3\mathrm{m},~\max_{j \in \mathcal{N}_{i}^{\mathrm{out}}}{\Delta_{i,j}}=16\mathrm{m}$, and the other parameters are the same as those in \figref{fig:CapacityCurves}.
The feasible set of data rate is $\mathcal{F}=\{ 5.12\mathrm{kbps} \}$.
}
\label{fig:IllustrationOfDataRateRange}
\end{figure}

An illustration is given in \figref{fig:IllustrationOfDataRateRange}.
\thmref{thm:feasible range of data rate} and \corref{cor:data rate without controller uncertainty} provide solutions to \probref{prob:First} in \secref{sec:Problem Description}.
With the fundamental limits of data rate, we can predetermine a proper data rate for transmission.
In \secref{sec:Closed-Loop Design of Formation Control and Communication Transmission}, for the transition process, we put forward an integrated design of formation control and power control to solve \probref{prob:Second}.

\section{Integrated Design of Formation Control and Power Control}\label{sec:Closed-Loop Design of Formation Control and Communication Transmission}

In this section, we propose an integrated design of formation control and power control for MASs under the data rate constraint to achieve bounded formation stability.
More specifically, \secref{sec:Formation Controller Design} gives the design of the formation control law and proves the bounded formation stability of MASs; \secref{sec:Integrated Design} first provides the power control strategy, and then gives an integrated design for MASs considering the constraints of the control layer and communication physical layer jointly; \secref{sec:Simulation} verifies the effectiveness of the proposed integrated design through numerical examples.

\subsection{Formation Control Design}\label{sec:Formation Controller Design}
Denote $\check{\mathbf{p}}_{i,k}=\mathbf{p}_{i,k}-\Delta_{i}$, $\mathbf{x}_{i,k}=[\check{\mathbf{p}}^{T}_{i,k},\mathbf{v}_{i,k}^{T}]^T$, $\mathbf{w}_{i,k}=[{\mathbf{w}_{i,k}^{p}}^T,{\mathbf{w}_{i,k}^{v}}^T]^T$ as the transformed position, the state, and the process noise with their realizations $\check{p}_{i,k} \in \llbracket \check{\mathbf{p}}_{i,k} \rrbracket \subseteq \mathbb{R}^n$, $x_{i,k} \in \llbracket\mathbf{x}_{i,k}\rrbracket \subseteq \mathbb{R}^{2n}$, and $w_{i,k} \in \llbracket\mathbf{w}_{i,k}\rrbracket \subseteq \mathbb{R}^{2n}$.
The dynamics of $i$\textsuperscript{th} agent can be rewritten as follows\footnote{In practical engineering, this discrete-time second-order model is commonly utilized to describe the dynamics of a vehicle, such as UAV~\cite{Wang2020,Wu2021}.}
\begin{align}\label{eqn:State Space Model}
\mathbf{x}_{i,k+1} = 
A\mathbf{x}_{i,k} 
+
B\mathbf{u}_{i,k}
+
\mathbf{w}_{i,k},
\end{align}
where $A=\begin{bmatrix} 1 & h \\ 0 & 1 \end{bmatrix} \otimes I_n,~B=\begin{bmatrix} \frac{h^2}{2} \\ h \end{bmatrix} \otimes I_n$.
The condition for achieving desired formation pattern \eqref{eqn:formation pattern achievement} is equivalent to $\lim_{k \to \infty} \| \mathbf{x}_{i,k}-\mathbf{x}_{j,k}   \| = 0$ for~$\forall i,j \in \mathcal{V}$.

Recalling the state information of neighbors cannot be instantly acquired, in this work, we consider the following estimation-based control protocol
\begin{align}\label{eqn:Modified Control Protocol}
u_{i,k}&=K\sum_{j \in \mathcal{N}_{i,c}}o_{ij}a_{ij}(\hat{x}^{[i]}_{j,k}-\hat{x}^{[i]}_{i,k}),
\end{align} 
where $K=[\alpha \quad \beta] \otimes I_n$ ($\alpha, \beta > 0$) is the control gain to be designed; $\mathcal{N}_{i,c}$ and $o_{ij}a_{ij}$ are the neighbor set of agent $i$ and the weight of edge $(j,i)$ over an undirected subgraph of the communication topology, respectively, which can be called the control topology $\mathcal{G}_c$.
More rigorous definition of control topology is defined as follows.

\begin{definition}[Control Topology]\label{def:control topology}
The control topology $\mathcal{G}_c=(\mathcal{V}_c,\mathcal{E}_c,\mathcal{A}_c)$ is an undirected subgraph of the communication topology $\mathcal{G}(m)=(\mathcal{V},\mathcal{E}(m),\mathcal{A}(m))$, where~$m=\lfloor k/\tau \rfloor \in \mathbb{N}_0$, satisfying $\mathcal{V}_c = \mathcal{V}$, $\mathcal{E}_c \subseteq \mathcal{E}(m)$, and $\mathcal{A}_c=[o_{ij}a_{ij}]$, where $o_{ij}=1$ iff $(j,i) \in \mathcal{E}_c$; otherwise, $o_{ij}=0$.
\end{definition}

Although the communication topology is time-varying, the control topology given in~\defref{def:control topology} is time-invariant (see our design of power control strategy in~\thmref{thm:Connectivity preservation of control topology}).
In addition, the sets of in-neighbors and out-neighbors for control topology~$\mathcal{G}_c$ are equal, which can be expressed by~$\mathcal{N}_{i,c}$.

In \eqref{eqn:Modified Control Protocol}, $\hat{x}^{[i]}_{l,k}$ ($l \in \mathcal{V}$) is the state estimate of $x_{l,k}$ from the side of agent $i$, with the following form
\begin{align}\label{eqn:State Estimation Update Law}
\hat{x}^{[i]}_{l,k}
=
\begin{cases}
A^kx_{l,0} & 1 \leq k \leq \tau-1, \\
A^{k-(\lfloor k/\tau \rfloor-1)\tau}x_{l,(\lfloor k/\tau \rfloor-1)\tau} & \tau \leq k,
\end{cases}
\end{align}
where for $l=i$ and $l \in \mathcal{V}\backslash \{i\}$, the available information for state estimate depends on $\mathcal{I}^{i}_{i,k}$ and $\mathcal{I}_{\mathcal{N}_{i,c},k}^i$, respectively.
It should be noted that the estimates for agent $l$ from any neighbor agent $i \in \mathcal{N}_{l,c}$ are the same.
Then, we define $\hat{x}_{l,k}:=\hat{x}^{[n_l]}_{l,k}$ for $\forall n_l \in \mathcal{N}_{l,c}$.

\begin{remark}
According to \eqref{eqn:AISs of in-neighbors}, for agent $i$ at time $k$, the latest available information of agent $j \in \mathcal{N}_{i,c}$ is $x_{j,(\lfloor k/\tau \rfloor-1)\tau}$.
Therefore, agent $i$ needs to estimate the states of neighbors for $k-(\lfloor k/\tau \rfloor-1)\tau$ sampling steps.
In addition, even though agent $i$ can obtain the accurate state of its own, it still utilizes the estimation in \eqref{eqn:Modified Control Protocol} to make the control inputs of its own predictable (see~\thmref{thm:Connectivity preservation of control topology}).
\end{remark}

\begin{remark}\label{rek:Controller uncertainty eliminate}
Since the state estimate is made at every sampling instant, the range of the control input is a ball of diameter $0$, which indicates that the proposed control protocol \eqref{eqn:Modified Control Protocol}-\eqref{eqn:State Estimation Update Law} can eliminate the uncertainty introduced by the controller (mentioned in \secref{sec:Analysis of Fundamental Limit for Data Rate}).
\end{remark}

To achieve bounded formation stability, the following theorem provides a sufficient condition for MASs under the constraint of a given data rate.

\begin{theorem}[Bounded Formation Stability]\label{thm:Formation Stability Theorem}

For the discrete-time MASs \eqref{eqn:State Space Model} with control protocol \eqref{eqn:Modified Control Protocol} and data rate $\mu$ (the corresponding transmission time is $T=\tau h$), the bounded formation stability (see~\defref{def:Bounded formation stability}) is achieved if:

\!\!\!\!\!\!(i) the control topology $\mathcal{G}_c$ is connected for $k \geq 0$;

\!\!\!\!\!\!(ii) the control gain satisfies $(\alpha,\beta) \in \mathcal{K}$, where
\!\!\!\!\!\!\begin{align}\label{eqn:control gain-1}
\mathcal{K}&=\bigcap_{i=2}^{N}\left\{ (\alpha,\beta)\colon\varphi_{1,i}>0,\varphi_{2,i}>0,\varphi_{3,i}>0 \right\}, \\
\varphi_{1,i}&=\bar{a}_{4}(i)-|\bar{a}_{0}(i)|,\\
\varphi_{2,i}&=|\bar{a}_{0}(i)^2-\bar{a}_{4}(i)^2|-|\bar{a}_{0}(i)\bar{a}_{3}(i)-\bar{a}_{1}(i)\bar{a}_{4}(i)|,\\
\nonumber
\varphi_{3,i}&=|(\bar{a}_{0}(i)^2-\bar{a}_{4}(i)^2)^2-(\bar{a}_{0}(i)\bar{a}_{3}(i)-\bar{a}_{1}(i)\bar{a}_{4}(i))^2| \\
\nonumber
&-|-(\bar{a}_{0}(i)\bar{a}_{1}(i)-\bar{a}_{3}(i)\bar{a}_{4}(i))(\bar{a}_{0}(i)\bar{a}_{3}(i)-\bar{a}_{1}(i)\bar{a}_{4}(i)) \\ \label{eqn:control gain-4}
&+\bar{a}_{2}(i)(\bar{a}_{0}(i)+\bar{a}_{4}(i))(\bar{a}_{0}(i)-\bar{a}_{4}(i))^2|,
\end{align}
$\bar{a}_{0}(i)=\lambda_i^2\alpha^2h^4\tau^2(\tau^2-1)/12,~\bar{a}_{1}(i)=\lambda_i \alpha h^2(\tau-2\tau^2)/2-\lambda_i \beta h\tau,~\bar{a}_{2}(i)=1+\lambda_i \alpha h^2(4\tau^2-\tau)/2+\lambda_i \beta h\tau,~\bar{a}_{3}(i)=-2$, and $\bar{a}_{4}(i)=1$.
\end{theorem}

\begin{IEEEproof}
See \apxref{apx:Proof of thm:Formation Stability Theorem}.
\end{IEEEproof}

\begin{remark}
    With~\eqref{eqn:control gain-1}-\eqref{eqn:control gain-4}, we calculate the feasible region of control gain through numerical simulations.
    The control gain~$(\alpha,\beta)$ is carefully selected based on this region.
\end{remark}

Although \thmref{thm:Formation Stability Theorem} provides a distributed formation controller for MASs to achieve bounded formation stability, the communication physical layer is still regarded as a black box.
In \secref{sec:Integrated Design}, we give an integrated design by jointly considering the constraints of formation control and communication transmission, where the connectivity condition of the control topology can be ensured for~$\forall k \in \mathbb{N}_0$.

\subsection{Integrated Design}\label{sec:Integrated Design}

Before designing the power control strategy, we define the function of states from neighbors in \eqref{eqn:Message} for $k=m\tau$ as
\begin{equation}\label{eqn:Message Content}
f_i(p_{\mathcal{N}_{i,c},k},v_{\mathcal{N}_{i,c},k}): =\sum_{j \in \mathcal{N}_{i,c}}{o_{ij}a_{ij}(\hat{x}_{j,k}-\hat{x}_{i,k})}.
\end{equation}

Then, with the SNR condition for successful transmission in \eqref{eqn:SNR Condition}, we provide a distributed power control strategy such that the connectivity of control topology can be guaranteed.

\begin{theorem}[Preservation of Control Topology]\label{thm:Connectivity preservation of control topology}
For the discrete-time MASs \eqref{eqn:State Space Model} with control protocol \eqref{eqn:Modified Control Protocol} and  data rate $\mu$, the initial connected control topology $\mathcal{G}_c$ preserves connectivity for~$k>0$ if the transmit power of agent $i$ satisfies
\begin{equation}\label{eqn:Transmit power update}
 P_{i,k}^{\mathrm{tx}} = \frac{(2^{\frac{\mu}{B_w}}-1)W}{g_{d_0}}(\frac{\bar{R}_{i,m\tau}}{d_0})^{\psi} + \epsilon,~~m=\lfloor \frac{k}{\tau_p} \rfloor,
\end{equation}
where~$\bar{R}_{i,m\tau}=\max_{j \in \mathcal{N}_{i,c}}\max_{l \in \{ 0,\ldots, \tau-1\}} \tilde{R}_{i,m\tau+l},~\tilde{R}_{i,k}=\|c_p(i,k)-c_p(j,k)\|+r_p(i,k)+r_p(j,k)$;~$\epsilon \geq 0$ is a bounded constant;~$W$ is the noise power of each agent independent of time~$k$.
The center and the radius of~$\llbracket \mathbf{p}_{s,k} \rrbracket^{[i]}=\mathcal{B}[c_p(s,k),r_p(s,k)]$,~$s \in \mathcal{N}_{i,c}\cup \{i\}$ are
\begin{align}
	c_p(s,k)&=p_{s,k-\sigma}+\sigma hv_{s,k-\sigma}+h^2\!\sum_{l=1}^{\sigma} (\sigma+\frac{1}{2}-l)u_{s,k-\sigma+l-1}, \\
    r_p(s,k)&=\sigma r_{w_p}(s)+\frac{1}{2}\sigma(\sigma-1)hr_{w_v}(s),
\end{align}
for~$\sigma>0$, and~$c(s,k)=p_{s,k},~r(s,k)=0$ for~$\sigma=0$.
Specifically, for agent~$i$ at time~$k \geq \tau$,~$\sigma=k-m\tau$; for agent~$j \in \mathcal{N}_{i,c}$ at time~$k \geq \tau$, $\sigma=k-(m-1)\tau$; for agent~$s \in \mathcal{N}_{i,c}\cup \{i\}$ at time~$0 \leq k < \tau$,~$\sigma=k$.
The predicted control input of agents~$i$ and~$j$ are
\begin{equation}
u_{i,m\tau+s_1}=KA^{\tau+s_1}\!\!\sum_{j \in \mathcal{N}_{i,c}}o_{ij}a_{ij}(x_{j,(m-1)\tau}-x_{i,(m-1)\tau}),
\end{equation}
\begin{equation}
\!\!\!\!u_{j,(m-1)\tau+s_2}=KA^{s_2} \!\!\!\!\!\!\sum_{j_1 \in \mathcal{N}_{j,c}} \!\!\!\!o_{jj_1}a_{jj_1}(\hat{x}_{j_1,(m-1)\tau}\!-\!\hat{x}_{j,(m-1)\tau}),
\end{equation}
\\
where~$s_1 \in \{ 0,\ldots, \tau-2\},~s_2 \in \{ 0,\ldots, 2\tau-2\}$.
\end{theorem}

\begin{IEEEproof}
	See \apxref{apx:Proof of thm:Connectivity preservation of control topology}.
\end{IEEEproof}

\begin{figure*}[!t]

\end{figure*}

Let~$\bar{R}^{\max}_{m\tau}=\max_{i,j \in \mathcal{V}}\max_{l \in \{ 0,\ldots, \tau-1\}} \|p_{i,m\tau+l}-p_{j,m\tau+l}\|$ be the maximum predicted distance between any pair of agents, and we get an upper bound of transmit power as follows
\begin{equation}
    P_{i,k}^{\mathrm{tx}} \leq \frac{(2^{\frac{\mu}{B_w}}-1)W}{g_{d_0}}(\frac{\bar{R}^{\max}_{m\tau}}{d_0})^{\psi}.
\end{equation}

\thmref{thm:Connectivity preservation of control topology} puts forward a power control strategy to enusre the connectivity of control topology combining the dynamics and the real-time communication model of agents, in which for each transmitter agent, the guaranteed communication radius takes into account the worst-case where the predicted distance between a pair of transmitter-receiver is the maximum.

Note that the power control strategy requires that the communication links between a pair of agents are bidirectional, which makes the control inputs of out-neighbors predictable.
Based on the existing framework in~\thmref{thm:Connectivity preservation of control topology}, it is difficult to generalize the power control strategy to directed and time-varying topologies.
In contrast, the estimation-based controller in~\thmref{thm:Formation Stability Theorem} can be easily extended to directed~\cite{Chen2024} and time-varying topology.

Then, with \thmref{thm:Formation Stability Theorem} and \thmref{thm:Connectivity preservation of control topology}, we provide a joint design in \thmref{thm:Closed-loop Design}.

\begin{theorem}[Integrated Design]\label{thm:Closed-loop Design}

For the discrete-time MASs \eqref{eqn:State Space Model} with control protocol \eqref{eqn:Modified Control Protocol} and  data rate $\mu$, the bounded formation stability can be achieved if the control gain satisfies $(\alpha,\beta) \in \mathcal{K}$ obtained in \thmref{thm:Formation Stability Theorem} and the transmit power control is executed based on \thmref{thm:Connectivity preservation of control topology}.
\end{theorem}

\begin{IEEEproof}
According to \thmref{thm:Formation Stability Theorem} and \thmref{thm:Connectivity preservation of control topology}, the control topology $\mathcal{G}_c$ maintains connected and the discrete-time MASs achieve bounded formation stability for $\forall k \in \mathbb{N}_0$.\hfill
\end{IEEEproof}

\thmref{thm:Closed-loop Design} provides an integrated design from the perspective of control and communication, in which the distributed control protocol is presented based on the real-time communication model, and the connectivity of the desired control topology is preserved according to the distributed power control strategy.
It is significantly different from the traditional control theory using simplified model of the communication links and can adapt to the time-varying communication conditions.\footnote{Note that our integrated design can largely reduce the bandwidth requirements by: (i) decreasing the data rate and (ii) increasing the transmit power.}

\begin{remark}
    With the communication model in~\secref{sec:System Model}, our integrated design provides a new perspective to study the formation control problem under the communication constraints.
    Compared to~\cite{9162475} and~\cite{Fan2024}, which focused on controlled movements, we not only analyze the channel capacity characteristics between controlled agents, but also reveal the relationship between control performance and channel capacity through the communication links.
\end{remark}

\begin{remark}
    In this work, the robustness of MASs affected by bounded process noises can be regarded as the ability to maintain bounded formation under a given control accuracy.
    With the proposed integrated design, the bounded formation of MASs with a given control accuracy can be guaranteed.
    Due to the bounded noises, there exists an upper limit of formation error even for systems without communication constraints.
    Thus, as long as the formation error is not less than the limit of that without communication constraints, the bounded process noises do not affect the robustness of MASs.
\end{remark}

\subsection{Numerical Examples}\label{sec:Simulation}

To demonstrate the effectiveness of the integrated design, a six-UAV system moving in a 2-D plane is considered.
The desired formation pattern is a triangle with lengths~$40\mathrm{m},~20\sqrt{2}\mathrm{m},~20\sqrt{2}\mathrm{m}$, and the edges in the control topology are with $o_{12} = o_{21} = o_{16} = o_{61} = o_{23} = o_{32} = o_{34} = o_{43} = o_{45} = o_{54} = o_{56} = o_{65} = 1$.
The position formation error bound is~$\delta_p=3\mathrm{m}$, the radii of position/velocity noise ranges for all UAVs are~$0.15\mathrm{m}$,~$\epsilon=0.0001\mathrm{w}$, and the other parameters are the same as those in \figref{fig:CapacityCurves}.

\begin{figure*}
\centering
\subfigure[] {\includegraphics [width=0.62\columnwidth]{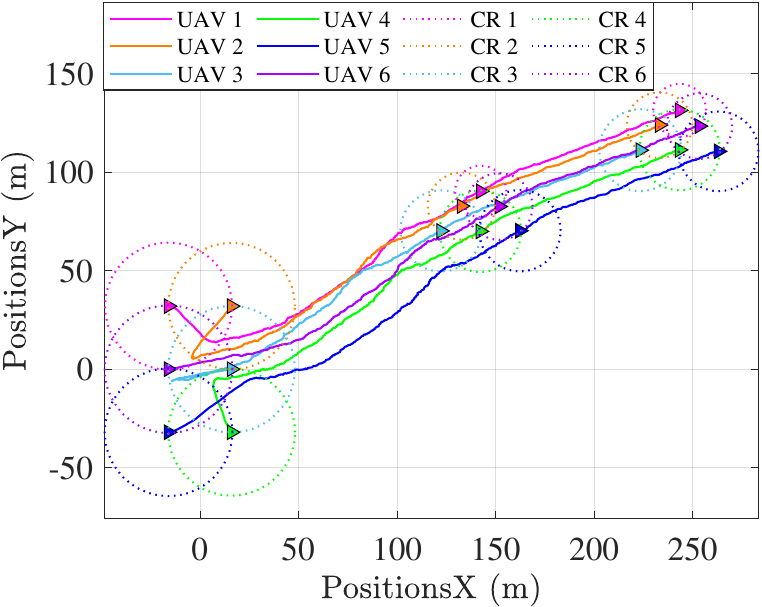}\label{fig:FormationPattern}} 
\subfigure[] {\includegraphics [width=0.653\columnwidth]{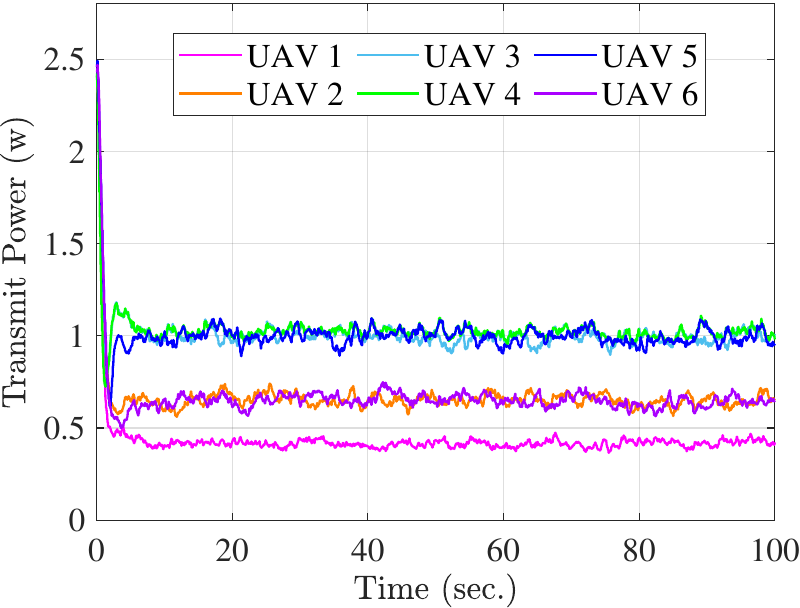}\label{fig:TransmitPowerControl}}
\subfigure[] {\includegraphics [width=0.65\columnwidth]{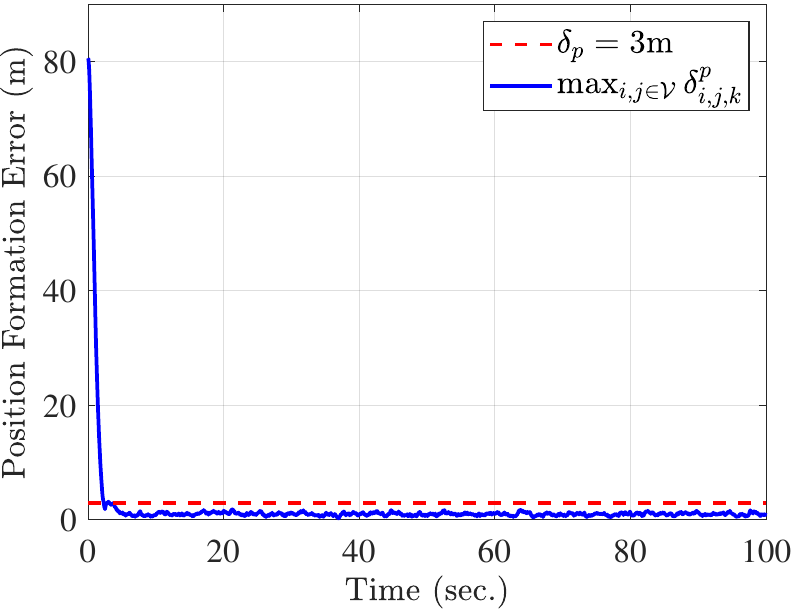}\label{fig:PositionFormationError}}
\caption{
    An illustrative example of formation control.
    (a) The position trajectories and communication regions (CRs).
    (b) Curves of the transmit power.
    (c) Curves of maximum formation error of position.
}
\label{fig:Formation}
\end{figure*}

With \corref{cor:data rate without controller uncertainty}, we choose the data rate as $\mu=5.12\mathrm{kbps}$, which means~$\tau=2$.
Then, based on~\thmref{thm:Closed-loop Design}, we select a feasible control gain $K=[1.54~1.61] \otimes I_2$.
As shown in \figref{fig:Formation}, with the integrated design, the six UAV systems will eventually form the desired regular triangle with bounded errors less than the required control accuracy, where the control topology preserves connectivity by modifying the transmit power.

\begin{figure*}
\centering
\subfigure[] {\includegraphics [width=0.65\columnwidth]{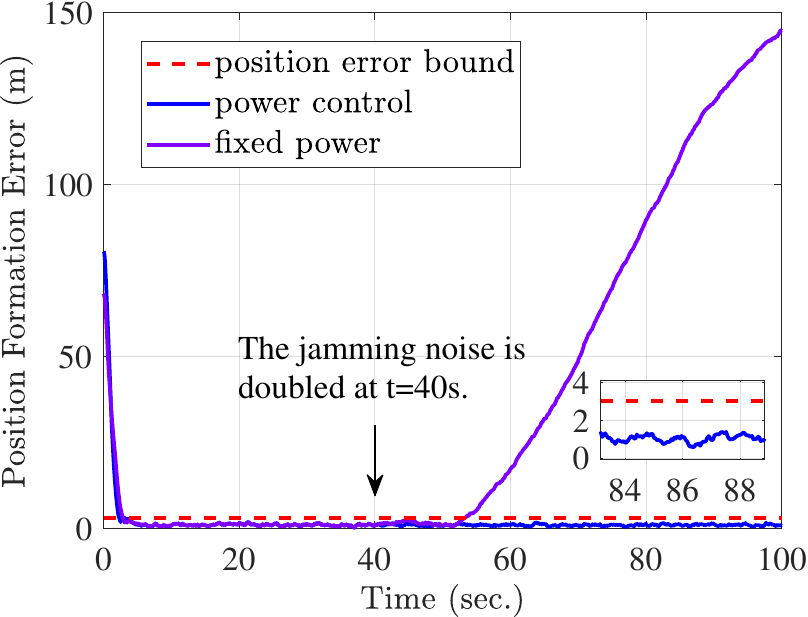}\label{fig:FormationErrorForChangedSNR}} 
\subfigure[] {\includegraphics [width=0.645\columnwidth]{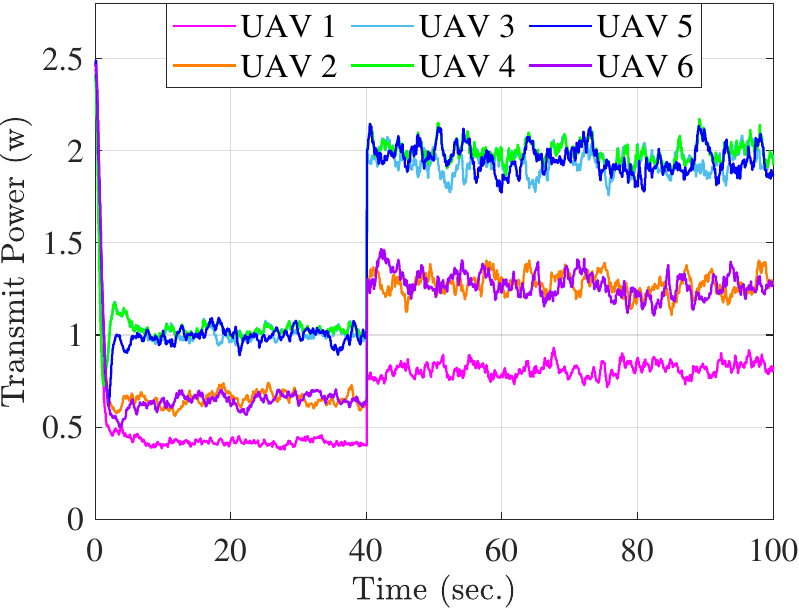}\label{fig:TransmitPowerModifiedSNR}}
\caption{
    A comparative example of formation control.
    (a) Curves of maximum formation error of position for the six UAV systems with adjustable and fixed transmit power, respectively, where the jamming noise is doubled at $t=40\mathrm{s}$.
    (b) Curves of the transmit power for UAVs with power control.
}
\label{fig:FormationPatternBreak}
\end{figure*}

For comparison, we simulate the formation control of the six UAV systems with fixed transmit power~$1.3452\mathrm{w}$.
Suppose that at $t=40\mathrm{s}$, the jamming noise is doubled to represent the time-varying communication condition.
In \figref{fig:FormationErrorForChangedSNR}, we give the curves of position formation error corresponding to our proposed power control strategy and fixed transmit power.\footnote{For the fixed transmit power case, the initial positions for UAVs are more compact than that in the power control case in order for the initial connectivity of communication topology.}
We can see that the system using power control strategy can achieve the required control accuracy.
The transmit power curves of UAVs are given in \figref{fig:TransmitPowerModifiedSNR}.

\begin{figure}
\centering
\includegraphics [width=0.7\columnwidth]{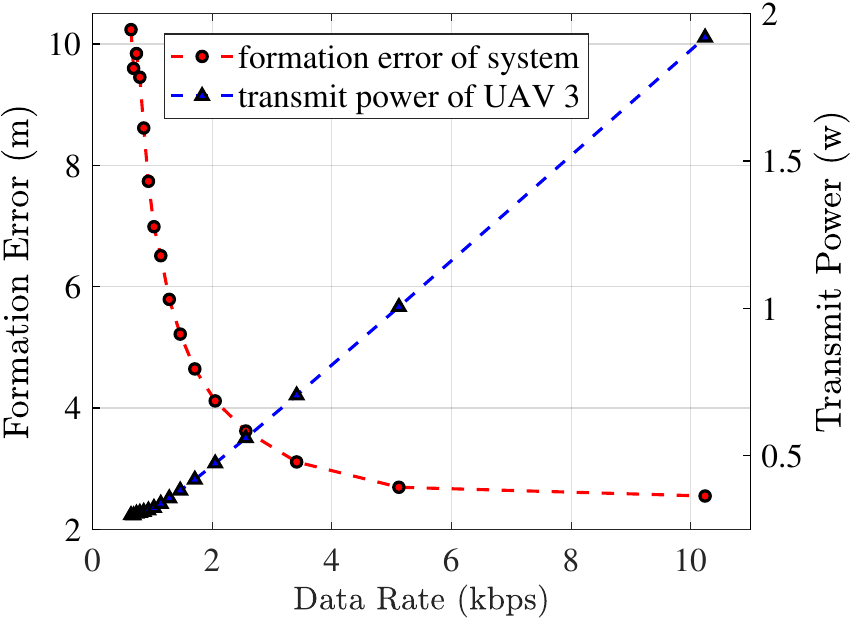}
\caption{
    Curves of (averaged) formation error for the six UAV systems and (averaged) transmit power for UAV~3 w.r.t. data rate in $100\mathrm{s}$, where the number of Monte Carlo experiments is set to 1000.
}
\label{fig:FEAndTPWithDataRate}
\end{figure}

In \figref{fig:FEAndTPWithDataRate}, we carried out Monte Carlo experiments to characterize the relationship of (averaged) formation error for the whole system, (averaged) transmit power for UAV~$3$, and data rate.
As the data rate grows, the formation error decreases and the transmit power increases.
For a high data rate, a single UAV must use a larger transmit power (which improves channel capacity) to accomplish the successful decoding of messages.
However, for a low data rate, if the noise ranges enlarge, the position uncertainty of each UAV increases according to~\secref{sec:Fundamental Limit for Data Rate}, which also requires a larger transmit power to expand the guaranteed communication region.

\section{Conclusion}\label{sec:Conclusion And Future Work}

In this article, we have studied the formation control problem of second-order MASs with bounded process noises, each agent using a unique communication band.
From the joint perspective of control and communication, a new concept has been introduced, called guaranteed communication region, to establish the model of communication links between agents.
The guaranteed communication region of a transmitter agent characterizes all possible locations for successful message decoding, which is determined by data rate, communication bandwidth, transmit power, power of noise, and location uncertainty of the transmitter agent.
Then, with the explicit expression of the position range, we have rigorously proved the concave property of the GCR w.r.t. the transmission time.
With this property, the fundamental limits of data rate for any desired formation accuracy have been obtained.
Furthermore, we have provided an integrated design of control and communication for the MASs with the desired data rate and formation accuracy, where the estimation-based controller and transmit power control strategy are proposed.

For future work, we will consider the communication interferences generated when multiple agents broadcast using the same bandwidth, and investigate how to balance the interferences and control performance.
Moreover, we can generalize the estimation-based controller and the power control strategy to directed and time-varying topologies for improving the applicability of our integrated design.

\appendices

\section{Proof of \propref{prop:Estimation of Uncertain Position Range}}\label{apx:Proof of prop:Estimation of Uncertain Position Range}
According to the prediction step of Corollary~1 in~\cite{9437974}, for each agent $i \in \mathcal{V}$ with dynamics \eqref{eqn:Second-Order of Position and Velocity}, we have:~$\llbracket\mathbf{p}_{i,k}\rrbracket^{[i]}=\llbracket\mathbf{p}_{i,k}|\mathcal{I}^i_{i,k} \rrbracket =p_{i,k}$, and
\begin{align}
\nonumber
\llbracket\mathbf{p}_{i,k}\rrbracket^{[i]} &=\llbracket\mathbf{p}_{i,k}|\mathcal{I}^i_{i,k-\sigma} \rrbracket \\ 
\nonumber
&=\left\{ p_{i,k-\sigma} \right\} \oplus \sigma h\left\{ v_{i,k-\sigma} \right\} \oplus \sum_{l=1}^{\sigma}\llbracket \mathbf{w}_{i,k-\sigma+l-1}^p \rrbracket \\
\nonumber
&\oplus h^2\sum_{l=1}^{\sigma}(\sigma+\frac{1}{2}-l)\llbracket \mathbf{u}_{i,k-\sigma+l-1} \rrbracket \\ \label{eqn:estimation of position range}
& \oplus h\sum_{l=1}^{\sigma} (\sigma-l)\llbracket \mathbf{w}_{i,k-\sigma+l-1}^v \rrbracket,
\end{align}
where $\sigma=k-\lfloor k/\tau \rfloor \tau$ and~$\sigma>0$.

Then, we give the following lemma to define the Minkowski sum of closed balls.

\begin{lemma}[Minkowski Sum of Closed Balls]\label{lem:Minkowski sum of balls}
For balls $\mathcal{B}[c_1,r_1], \ldots, \mathcal{B}[c_n,r_n] \in \mathbb{R}^n$, 
\begin{align}\label{eqn:Ball Operation}
\mathcal{B}[c_{1:n},r_{1:n}]=\mathcal{B}[c_1,r_1] \oplus \cdots \oplus \mathcal{B}[c_n,r_n],
\end{align}
where $c_{1:n}=c_1+\cdots+c_n$, $r_{1:n}=r_1+\cdots+r_n$.
\end{lemma}

With \lemref{lem:Minkowski sum of balls}, the position range given in~\eqref{eqn:estimation of position range} is an $n$-dimensional closed ball, i.e., $\llbracket\mathbf{p}_{i,k}\rrbracket^{[i]}=\mathcal{B}[c_p(i,k),r_p(i,k)] \in \mathbb{R}^n$.
Then, the center $c_p(i,k)$ and the radius $r_p(i,k)$ are
\begin{align}
\nonumber
c_p(i,k)&=p_{i,k-\sigma}+\sigma hv_{i,k-\sigma} \\
& +h^2\sum_{l=1}^{\sigma}(\sigma+\frac{1}{2}-l)c_u(i,k-\sigma+l-1), \\
\nonumber
r_p(i,k)&=h^2\sum_{l=1}^{\sigma}(\tau+\frac{1}{2}-l)r_u(i,k-\tau+l-1) \\
&+\sigma r_{w_p}(i)+\frac{1}{2}\sigma(\sigma-1)hr_{w_p}(i),
\end{align}
for~$\sigma \in \{ 1,\ldots,\tau-1 \}$.
Specially, for~$\sigma=0$,~$c_p(i,k)=p_{i,k},~r_p(i,k)=0$.\hfill$\blacksquare$

\section{Proof of \lemref{lem:Guaranteed Communication Radius}}\label{apx:Proof of lem:Guaranteed Communication Radius}
In the following analysis, the indices~$i$~and $k$ are neglected for convenience.
With~\defref{def:Communication Range of Point}, the communication region of agent~$i$ located at~$p \in \mathcal{B}[c_p,r_p]$ is an~$n$-dimensional open ball~${\mathcal{B}}(p,R)$, where~$R$ is given in~\eqref{eqn:Communication Radius}.
Therefore, for any transmitter agent~$p_{\mathrm{tx}} \in \mathcal{B}[c_p,r_p]$ and receiver agent~$p_{\mathrm{rx}} \in \Omega_{\llbracket\mathbf{p}_{i,k}\rrbracket^{[i]}}$, the distance should satisfy:~$\|p_{\mathrm{tx}}-p_{\mathrm{rx}}\| < R$.
Based on~\defref{def:Communication Range With Uncertain}, we prove that the guaranteed communication region~$\Omega_{\llbracket\mathbf{p}_{i,k}\rrbracket^{[i]}}$ is an~$n$-dimensional open ball~${\mathcal{B}}(c_p,R-r_p)$ by the following two steps:

\!\!\!\!\!(i) $\forall p_{\mathrm{tx}} \in \mathcal{B}[c_p,r_p],~p_{\mathrm{rx}} \in {\mathcal{B}}(c_p,R-r_p)$, we have~$\|p_{\mathrm{tx}}-c_p\| \leq r$ and~$\|c_p-p_{\mathrm{rx}}\|<R-r_p$.
With~$\|p_{\mathrm{tx}}-p_{\mathrm{rx}}\| \leq \|p_{\mathrm{tx}}-c_p\|+\|c_p-p_{\mathrm{rx}}\|$, we can get~$\|p_{\mathrm{tx}}-p_{\mathrm{rx}}\|<R$;

\!\!\!\!\!(ii) $\exists p_{\mathrm{tx}}=c_p+r_pe \in \mathcal{B}[c_p,r_p],~p_{\mathrm{rx}}=c_p-\rho_p e \in \mathcal{B}(c_p,\rho_p),~e \in \mathbb{R}^n,~\|e\|=1$, and~$\rho_p \geq R-r_p$, such that~$\|p_{\mathrm{tx}}-p_{\mathrm{rx}}\|=(\rho_p+r_p)\|e\| \geq R$, which violates~$\|p_{\mathrm{tx}}-p_{\mathrm{rx}}\|<R$.\hfill$\blacksquare$

\section{Proof of \propref{prop:Concave Property}}\label{apx:Proof of prop:Concave Property}
For a transmitter agent $i \in \mathcal{V}$ with dynamics \eqref{eqn:Second-Order of Position and Velocity} and $r_u(i,k)=0$, according to~\eqref{eqn:radius without control uncertainty}, the radius of position range $\llbracket\mathbf{p}_{i,k'}\rrbracket^{[i]} =\llbracket\mathbf{p}_{i,k'}|\mathcal{I}^i_{i,\lfloor k'/\tau \rfloor\tau}\rrbracket=\mathcal{B}[c_p(i,k'),r_p(i,k')]$ (which means $\sigma=\tau-1$) at time $k'=m\tau-1$~($m \in \mathbb{Z}_+$) is
\begin{align}
\nonumber
r_p(i,k')&=(\tau-1)r_{w_p}(i)+\frac{1}{2}(\tau-1)(\tau-2)hr_{w_v}(i).
\end{align}
The GCR for agent $i$ is~$R_{\llbracket\mathbf{p}_{i,k'}\rrbracket^{[i]}}=R_{i,k'}-r_p(i,k')$, where $R_{i,k'}$ is the communication radius of agent $i$ located at any position $p_{i,k'} \in {\mathcal{B}}[c_p(i,k'),r_p(i,k')]$.
Obviously, $r_p(i,k')$ is quadratic w.r.t. $\tau$, the same as~$-r_p(i,k')$.

For~$R_{i,k'}$, we take the first and second derivatives as follows
\begin{align}\label{eqn:g_1}
\frac{dR_{i,k'}}{d\tau}&=\frac{Md_0\ln2}{hB_w\psi}\left (\frac{g_{d_0}P_{i,k'}^{\mathrm{tx}}}{W_{j,k'}}\right )^{1/\psi}\frac{2^{\frac{M}{\tau hB_w}}}{\tau^2(2^{\frac{M}{\tau hB_w}}-1)^{1+\frac{1}{\psi}}}, \\
\frac{d^2R_{i,k'}}{d\tau^2}&=D_1(\tau)D_2(\tau),
\end{align}
where
\begin{align}
D_1(\tau)&=\frac{Md_0\ln2}{h^2B_w^2\psi^2}\left (\frac{g_{d_0}P_{i,k'}^{\mathrm{tx}}}{W_{j,k'}}\right )^{1/\psi}\frac{2^{\frac{M}{\tau hB_w}}}{\tau^4(2^{\frac{M}{\tau hB_w}}-1)^{2+\frac{1}{\psi}}}, \\
D_2(\tau)&=2\tau hB_w \psi \left ( 1-2^{\frac{M}{\tau hB_w}} \right )+M\ln2(\psi+2^{\frac{M}{\tau hB_w}}).
\end{align}
Since the parameters $M,~h,~B_w,~P_{i,k'}^{\mathrm{tx}},~W_{j,k'},~\psi$ are positive, it is easy to verify $dR_{i,k'}/d\tau>0$ and $D_1(\tau)>0$.
If $D_2(\tau)<0$, we have $d^2R_{i,k'}/d\tau^2=D_1(\tau)D_2(\tau) < 0$, which indicates $R_{i,k'}(\tau)$ is concave w.r.t. $\tau$. 
Thus, we can obtain that the GCR $R_{\llbracket\mathbf{p}_{i,k'}\rrbracket^{[i]}}=R_{i,k'}-r_p(i,k')$ is concave w.r.t. $\tau$.

In addition, with the first derivation of $R_{\llbracket\mathbf{p}_{i,k'}\rrbracket^{[i]}}$, we can always find a constant $\mathring{\tau}_{i,k'} \in \mathbb{Z}_+$ such that $dR_{\llbracket\mathbf{p}_{i,k'}\rrbracket^{[i]}}/d\tau=0$.
Therefore, $R_{\llbracket\mathbf{p}_{i,k'}\rrbracket^{[i]}}$ increases monotonically for $\tau \in (0,\mathring{\tau}_{i,k'}]$, and decreases monotonically for $\tau \in (\mathring{\tau}_{i,k'},\infty)$.\hfill$\blacksquare$

\section{Proof of \corref{cor:data rate without controller uncertainty}}\label{apx:Proof of cor:data rate without controller uncertainty}
With~\propref{prop:Concave Property} and~\rekref{rek:bandwidth limit}, if $r_u(i,k)=0,~\psi>1$, and $B_w>\underline{B}>0$, the concave property of $R_{\llbracket\mathbf{p}_{i,k'}\rrbracket^{[i]}}$ holds for time $k'=m\tau-1$.
With $r_u(i,k)=0,~r_u(j,k)=0~(j \in \mathcal{V}\backslash\{i\})$, and footnote~11, the radii of position ranges for all agents are predictable.
Specifically, from the side of agent $i$, the predict step satisfies:~$\sigma=\tau-1$ for itself and~$\sigma=2\tau-1$ for agent~$j \in \mathcal{V}\backslash\{i\}$.
Then, with \eqref{eqn:radius without control uncertainty}, we can obtain the explicit expressions of $\Psi_{i,k'}^1(\tau)$ and $\Psi_{i,k'}^2(\tau)$ given in \eqref{eqn:explicit form-1}-\eqref{eqn:explicit form-2}.
Since the lower and upper bounds of~$\tau$ are determined by~$\Psi_{i,k'}^1(\tau)$ and~$\Psi_{i,k'}^2(\tau)$, respectively, the feasible set of data rate is
\begin{equation}
\nonumber
\mathcal{F}_{i,k'}=\left\{\frac{M}{\tau h} \colon S^{\min}(\Psi_{i,k'}^1) \leq \tau \leq S^{\max}(\Psi_{i,k'}^2),~\tau \in \mathbb{Z}_+ \right\}.
\end{equation}\hfill$\blacksquare$

\section{Proof of \thmref{thm:Formation Stability Theorem}}\label{apx:Proof of thm:Formation Stability Theorem}
In this section, since the states and process noises for agents at each sampling instant are deterministic in practice, we use the realizations of corresponding uncertain variables for analysis.
The analysis results are valid for all realizations within the range of any uncertain variable.

With \eqref{eqn:State Estimation Update Law}, the state estimate of MASs, i.e.,~$\hat{x}_{k}:=[\hat{x}_{1,k}^T,\ldots,\hat{x}_{N,k}^T]^T$, can be compactly described by
\begin{equation}\label{eqn:Estimation of Whole System}
\!\!\!\!\hat{x}_k
=
\begin{cases}
(I_N \otimes A^{k})x_{0} & 1 \leq k \leq \tau-1 \\
(I_N \otimes A^{k-(m-1)\tau})x_{(m-1)\tau} & \tau \leq k
\end{cases},
\end{equation}
where $m=\lfloor k/\tau \rfloor \in \mathbb{Z}_+$.
Let ${x}_{k}=[{x}_{1,k}^T,\ldots,{x}_{N,k}^T]^T$ and~${w}_{k}=[{w}_{1,k}^{T},\ldots,{w}_{N,k}^{T}]^{T}$, the closed-loop system~(for $k \geq \tau$) can be described as follows
\begin{multline}\label{eqn:closed-loop}
{x}_{k+1}
=(I_N \otimes A){x}_{k}-\\(\mathcal{L}_c \otimes BKA^{k-(m-1)\tau})x_{(m-1)\tau}+{w}_{k},
\end{multline}
where $\mathcal{L}_c$ is the Laplacian matrix of the control topology~$\mathcal{G}_c$.

The formation error of the MASs is defined by
\begin{equation}
{\delta}_k:={x}_k-\mathbf{1}_N \otimes \bar{{x}}_k:=[{\delta}_{1,k}^T,\ldots,{\delta}_{N,k}^T ]^T,
\end{equation}
where $\bar{{x}}_k=(1/N)\sum_{i=1}^{N}{x}_{i,k}$ is the average state of all agents.
Then, the bounded formation stability of the MASs can be equivalently measured by the boundedness of~$\delta_k$.

With $\bar{{x}}_k=(1/N)(\mathbf{1}_N^T \otimes I_{2n}){x}_k$ and $\mathbf{1}_N^T\mathcal{L}_c=\mathbf{0}_N^T$, we have
\begin{align}\label{eqn:Average State}
\bar{{x}}_{k+1}=A\bar{{x}}_{k}+\bar{{w}}_k,
\end{align}
where $\bar{{w}}_k=(1/N)(\mathbf{1}_N^T \otimes I_{2n}){w}_k$.
Combined with ${\delta}_{k}={x}_k-\mathbf{1}_N \otimes \bar{{x}}_k$ and $\mathcal{L}_c\mathbf{1}_N=\mathbf{0}_N$, the error MASs can be described by
\begin{multline}\label{eqn:Compact Form of System With State Estimation}
{\delta}_{k+1}
=(I_N\otimes A){\delta}_k -\\
(\mathcal{L}_c \otimes BKA^{k-(m-1)\tau})\delta_{(m-1)\tau}+{\delta}_k^{w},
\end{multline}
where ${\delta}_k^w={w}_k-\mathbf{1}_N \otimes \bar{{w}}_k$.

Select $q_i \in \mathbb{R}^N$ such that $q_i^T\mathcal{L}_c=\lambda_iq_i^T$ and form an unitary orthogonal matrix $Q=\begin{bmatrix} \frac{1}{\sqrt{N}}\mathbf{1}_N & q_2 & \ldots & q_N \end{bmatrix} \in \mathbb{R}^{N \times N}$ to transform $\mathcal{L}_c$ into a diagonal form
\begin{align}\label{eqn:Diagonal Form of Laplacian Matrix}
\mathcal{J}=Q^T\mathcal{L}_cQ=\mathrm{diag}\{ \lambda_1,\lambda_2,\ldots,\lambda_N \},
\end{align}
where $\lambda_1,\ldots,\lambda_N$ are the eigenvalues of the Laplacian matrix $\mathcal{L}_c$.
If the control topology $\mathcal{G}_c$ is connected, we can obtain $\lambda_1=0$ and $\lambda_2,\ldots,\lambda_N>0$.

Denote ${\delta}_k=(Q \otimes I_{2n})\tilde{{\delta}}_k$ and ${\delta}_k^w=(Q \otimes I_{2n})\tilde{{\delta}}_k^w$, the error MASs \eqref{eqn:Compact Form of System With State Estimation} can be converted as follows
\begin{multline}\label{eqn:Linear Transformation of Kronecker Basis}
\tilde{{\delta}}_{k+1}=(I_N\otimes A)\tilde{{\delta}}_k- \\
(\mathcal{J} \otimes BKA^{k-(m-1)\tau})\tilde{\delta}_{(m-1)\tau}+\tilde{{\delta}}_k^{w}.
\end{multline}

Partition $\tilde{{\delta}}_k$ into $N$ components, i.e., $\tilde{{\delta}}_{k}:=[ \tilde{{\delta}}_{k,(1)}^T,\ldots,\tilde{{\delta}}_{k,(N)}^T]^T$, where $\tilde{{\delta}}_{k,(i)} \in \mathbb{R}^{2n},~i \in \{ 1,\ldots,N \}$.
The decomposition of $\tilde{{\delta}}_k^w$ is similar to $\tilde{{\delta}}_k$.
Thus, each component can be described by
\begin{equation}\label{eqn:consensus subspace of i}\!\!\!\!\!
\tilde{{\delta}}_{k+1,(i)}=A\tilde{{\delta}}_{k,(i)}-\lambda_iBKA^{k-(m-1)\tau}\tilde{\delta}_{(m-1)\tau,(i)}+\tilde{{\delta}}_{k,(i)}^w,
\end{equation}
Moreover, the error $\tilde{{\delta}}_{k+1,(i)}$ can be rewritten as follows:
\begin{align}\label{eqn:accumulation}
\nonumber
\tilde{{\delta}}_{k+1,(i)}&=A^{k-m\tau+1}\tilde{\delta}_{m\tau,(i)}-N_i(k-m\tau+1)\tilde{\delta}_{(m-1)\tau,(i)} \\
&+\sum_{l=0}^{k-m\tau}A^{k-m\tau-l}\tilde{{\delta}}_{m\tau+l,(i)}^w,
\end{align}
where 
\begin{align}\label{eqn:N_i}
\!\!N_i(k-m\tau+1) =\lambda_i \!\!\!\!\sum_{l=0}^{k-m\tau}\!\!\!\!{A^{k-m\tau-l}BKA^{k-m\tau+1+l}}.
\end{align}
Let $s=k-m\tau+1 \in \{ 1,\ldots,\tau \}$, and then substitute the matrices $A$ and $B$ into \eqref{eqn:N_i}:
\begin{equation}\!\!\!\!
N_i(s)=\lambda_i
\begin{bmatrix}
\frac{\alpha h^2s^2}{2} & \frac{\alpha h^3}{12}s (8s^2-3s+1)+\frac{\beta h^2s^2}{2} \\ \alpha hs & \frac{\alpha h^2}{2}s(3s-1)+\beta hs
\end{bmatrix}
\otimes I_n.
\end{equation}

Especially, with ${\delta}_k=(Q \otimes I_{2n})\tilde{{\delta}}_k$, we can immediately get
\begin{align}
\tilde{{\delta}}_{k,(1)}=(1/\sqrt{N}\mathbf{1}_N \otimes I_{2n})^T(x_k-\mathbf{1}_N \otimes \bar{x}_k)=\mathbf{0}_{2n}.
\end{align}

Define ${Z}_{(m+1)\tau,(i)}=\begin{bmatrix} \tilde{{\delta}}_{(m+1)\tau,(i)}^T & \tilde{{\delta}}_{m\tau,(i)}^T \end{bmatrix}^T$, we have
\begin{align}\label{eqn:Subsystem With Compact Matrix}
{Z}_{(m+1)\tau,(i)}&=\bar{A}_i(\tau){Z}_{m\tau,(i)}+\bar{{U}}_{m\tau,(i)},
\end{align}
where~$i \in \{ 2,\ldots,N \}$,
\begin{align}
\bar{A}_i(\tau) & =\begin{bmatrix} A^{\tau} & -N_i(\tau) \\ I_{2n} & \mathbf{0}_{2n \times 2n} \end{bmatrix}, \\
\bar{{U}}_{m\tau,(i)} & =\begin{bmatrix} \sum_{l=0}^{\tau-1}A^{\tau-1-l}\tilde{{\delta}}^w_{m\tau+l,(i)} \\ \mathbf{0}_{2n \times 1} \end{bmatrix}.
\end{align}

Let $n=1$, and the higher dimension can be generalized by Kronecker product.
Then, we should to ensure that $\bar{A}_i(\tau)$ is Schur stable.
The characteristic polynomial $f(z)$ corresponding to $\bar{A}_i(\tau)$ can be obtained by the following determinant 
\begin{align}\label{eqn:Determinant}
f(z)&=\mathrm{det}(zI_{4}-\bar{A}_i(\tau))=\mathrm{det}\begin{pmatrix}
\tilde{A}_{11} & \tilde{A}_{12} \\
\tilde{A}_{21} & \tilde{A}_{22}
\end{pmatrix},
\end{align}
where
\begin{align}
\nonumber
\tilde{A}_{11}=
\begin{bmatrix}
z-1 & -h\tau \\ 0 & z-1
\end{bmatrix},
\tilde{A}_{21}=
\begin{bmatrix}
-1 & 0 \\ 0 & -1
\end{bmatrix},
\tilde{A}_{22}=
\begin{bmatrix}
z & 0 \\ 0 & z
\end{bmatrix},
\end{align}
\begin{align}
\nonumber
\tilde{A}_{12}=
\begin{bmatrix}
-\lambda_i \frac{\alpha h^2}{2}\tau^2 &  -\lambda_i( \frac{\alpha h^3}{12}\tau (8\tau^2-3\tau+1)+\frac{\beta h^2}{2}\tau^2 ) \\ \lambda_i \alpha h\tau & \lambda_i(\frac{\alpha h^2}{2}\tau(3\tau-1)+\beta h\tau)
\end{bmatrix},
\end{align}
and we can readily get\footnote{For a block matrix $\tilde{A}=\begin{pmatrix} S_{11} & S_{12} \\ S_{21} & S_{22} \end{pmatrix}$, where $\tilde{A}_{11}, \tilde{A}_{12}, \tilde{A}_{21}, S_{22}$ are square matrices with the same dimension, if $\tilde{A}_{21}\tilde{A}_{22}=\tilde{A}_{22}\tilde{A}_{21}$, $\mathrm{det}(\tilde{A})=\mathrm{det}(\tilde{A}_{11}\tilde{A}_{22}-\tilde{A}_{12}\tilde{A}_{21})$.}
\begin{multline}\label{eqn:Determinant Based On Lemma}
f(z)=z^4-2z^3+z^2[1+\lambda_i \frac{\alpha h^2}{2}(4\tau^2-\tau)+\lambda_i \beta h\tau]+ \\
z[\lambda_i \frac{\alpha h^2}{2}(\tau-2\tau^2)-\lambda_i \beta h\tau]+\lambda_i^2\frac{\alpha^2h^4}{12}\tau^2(\tau^2-1).
\end{multline}
The parameters of Jury table can be constructed as
\begin{align}
\nonumber
\bar{b}_{0}(i)&=\begin{vmatrix} \bar{a}_{0}(i) & \bar{a}_{4}(i) \\ \bar{a}_{4}(i) & \bar{a}_{0}(i) \end{vmatrix}, \quad \bar{b}_{1}(i)=\begin{vmatrix} \bar{a}_{0}(i) & \bar{a}_{3}(i) \\ \bar{a}_{4}(i) & \bar{a}_{1}(i) \end{vmatrix}, \\
\nonumber
\bar{b}_{2}(i)&=\begin{vmatrix} \bar{a}_{0}(i) & \bar{a}_{2}(i) \\ \bar{a}_{4}(i) & \bar{a}_{2}(i) \end{vmatrix}, \quad \bar{b}_{3}(i)=\begin{vmatrix} \bar{a}_{0}(i) & \bar{a}_{1}(i) \\ \bar{a}_{4}(i) & \bar{a}_{3}(i) \end{vmatrix}, \\
\nonumber
\bar{c}_{0}(i)&=\begin{vmatrix} \bar{b}_{0}(i) & \bar{b}_{3}(i) \\ \bar{b}_{3}(i) & \bar{b}_{0}(i) \end{vmatrix}, \quad \bar{c}_{1}(i)=\begin{vmatrix} \bar{b}_{0}(i) & \bar{b}_{2}(i) \\ \bar{b}_{3}(i) & \bar{b}_{1}(i) \end{vmatrix}, \\
\nonumber
\bar{c}_{2}(i)&=\begin{vmatrix} \bar{b}_{0}(i) & \bar{b}_{1}(i) \\ \bar{b}_{3}(i) & \bar{b}_{2}(i) \end{vmatrix},
\end{align}
where
$\bar{a}_{0}(i)=\lambda_i^2\alpha^2h^4\tau^2(\tau^2-1)/12,~\bar{a}_{1}(i)=\lambda_i \alpha h^2(\tau-2\tau^2)/2-\lambda_i \beta h\tau,~\bar{a}_{2}(i)=1+\lambda_i \alpha h^2(4\tau^2-\tau)/2+\lambda_i \beta h\tau,~\bar{a}_{3}(i)=-2,~\bar{a}_{4}(i)=1$.

According to Jury stable criterion,~$\bar{A}_i(\tau)$ is Schur stable if and only if the following conditions hold
\begin{itemize}
\item $f(1)=\lambda_i \alpha h^2\tau^2+\lambda_i^2\frac{\alpha^2h^4}{12}\tau^2(\tau^2-1)>0$,
\item $(-1)^{4}f(-1)>0$,
\item $|\bar{a}_{0}(i)|<\bar{a}_{4}(i)$, $|\bar{b}_{3}(i)|<|\bar{b}_{0}(i)|$, $|\bar{c}_{2}(i)|<|\bar{c}_{0}(i)|$.	
\end{itemize}
It is easy to derive that $f(1)>0$ and $(-1)^{4}f(-1)>0$ hold for $\forall \tau \in \mathbb{Z}_+$ and~$i \in \{ 2,\ldots,N \}$.
Then, we define
\begin{align}
\varphi_{1,i}&=\bar{a}_{4}(i)-|\bar{a}_{0}(i)|,\\
\varphi_{2,i}&=|\bar{a}_{0}(i)^2-\bar{a}_{4}(i)^2|-|\bar{a}_{0}(i)\bar{a}_{3}(i)-\bar{a}_{1}(i)\bar{a}_{4}(i)|,\\
\nonumber
\varphi_{3,i}&=|(\bar{a}_{0}(i)^2-\bar{a}_{4}(i)^2)^2-(\bar{a}_{0}(i)\bar{a}_{3}(i)-\bar{a}_{1}(i)\bar{a}_{4}(i))^2| \\
\nonumber
&-|-(\bar{a}_{0}(i)\bar{a}_{1}(i)-\bar{a}_{3}(i)\bar{a}_{4}(i))(\bar{a}_{0}(i)\bar{a}_{3}(i)-\bar{a}_{1}(i)\bar{a}_{4}(i)) \\
&+\bar{a}_{2}(i)(\bar{a}_{0}(i)+\bar{a}_{4}(i))(\bar{a}_{0}(i)-\bar{a}_{4}(i))^2|.
\end{align}
If $\varphi_{1,i},~\varphi_{2,i},~\varphi_{3,i}>0$ for $\forall i \in \{ 2,\ldots,N \}$, we can always find feasible solutions of the control gain (i.e., $\alpha$ and $\beta$).
That is, if the control gain $K$ satisfies $(\alpha,\beta) \in \mathcal{K}$, where
\begin{align}
\mathcal{K}&=\bigcap_{i=2}^{N}\left\{ (\alpha,\beta)\colon\varphi_{1,i}>0,~\varphi_{2,i}>0,~\varphi_{3,i}>0 \right\},
\end{align}
the matrix $\bar{A}_i(\tau)$ is Schur stable.

For the noise term in \eqref{eqn:closed-loop} at time $k \in \mathbb{N}_0$, we have
\begin{align}
&\|{w}_k\| \leq r_w,
\end{align}
where
\begin{equation}
	\nonumber
	r_w =\sqrt{r_{w_p}(1)^2+r_{w_v}(1)^2+\cdots+r_{w_p}(N)^2+r_{w_p}(N)^2}.
\end{equation}

With $\tilde{{\delta}}^w_{k,(i)}=(q_i \otimes I_{2n})^T{\delta}^w_{k}$ and $\|(q_i \otimes I_{2n})^T\|=\|I_{2nN}-\frac{1}{N}\mathbf{1}_N \otimes (\mathbf{1}_N^T \otimes I_{2n})\|=1$, we can immediately obtain 
\begin{align}\label{eqn:88}
\|\tilde{{\delta}}^w_{k,(i)}\| \leq \|{w}_k\| + \|\mathbf{1}_N \otimes \bar{w}_k\| \leq 2\|{w}_k\| \leq 2r_w.
\end{align}
Then, the second term related to noise in \eqref{eqn:Subsystem With Compact Matrix} is bounded:
\begin{align}
\nonumber
\|\bar{U}_{m\tau,(i)}\|&=\left\|  \begin{bmatrix} \sum_{l=0}^{\tau-1}A^{\tau-1-l}\tilde{{\delta}}^w_{m\tau+l,(i)} \\ \mathbf{0}_{2n \times 1} \end{bmatrix} \right\| \\  \label{eqn:89}
&\leq 2\sum_{l=0}^{\tau-1}\left\| \begin{matrix}A^{\tau-1-l}\end{matrix} \right\| r_w.
\end{align}

Proceeding forward, for \eqref{eqn:Subsystem With Compact Matrix}, we have
\begin{equation}
{Z}_{(m+1)\tau,(i)} = \bar{A}_i(\tau)^m {Z}_{\tau,(i)}+\sum_{l=1}^{m}\bar{A}_i(\tau)^{m-l}\bar{{U}}_{l\tau,(i)}.
\end{equation}

Since the matrix $\bar{A}_i(\tau)$ is Schur stable when $(\alpha,\beta) \in \mathcal{K}$, i.e., the eigenvalues of $\bar{A}_i(\tau)$ are located strictly inside the unit disk, there are constants $c_i>0$ and $0 \leq \sigma_i <1$ such that $|\bar{A}_i(\tau)^m| \leq c_i\sigma^m_i$~\cite{Input-to-state_stability_for_discrete-time_nonlinear_systems}.
Thus, there exist a $\mathcal{K} \mathcal{L}$-function $\phi_1: \mathbb{R}_{\geq 0} \times \mathbb{R}_{\geq 0} \rightarrow \mathbb{R}_{\geq 0}$ and a $\mathcal{K}$-function $\phi_2: \mathbb{R}_{\geq 0} \rightarrow \mathbb{R}_{\geq 0}$ such that, for each bounded noise term $\bar{{U}}_{l\tau,(i)} \in \mathbb{R}^{4n \times 1}$ and initial state ${Z}_{\tau,(i)} \in \mathbb{R}^{4n \times 1}$, it holds that 
\begin{align}\label{eqn:ISS-1}
\|{Z}_{(m+1)\tau,(i)}\| \leq \phi_1(\|{Z}_{\tau,(i)}\|)+\phi_2(\|\bar{{U}}_{l\tau,(i)}\|),
\end{align}
where
\begin{align}
\nonumber
\phi_1(\|{Z}_{\tau,(i)}\|)&=c_i\|{Z}_{\tau,(i)}\|, \\
\nonumber
\phi_2(\|\bar{{U}}_{l\tau,(i)}\|)&=\sum_{l=0}^{\infty}c_i\sigma^l_i \|\bar{{U}}_{l\tau,(i)}\|=\frac{c_i \|\bar{{U}}_{l\tau,(i)}\|}{1-\sigma_i}.
\end{align}
Therefore, system \eqref{eqn:Subsystem With Compact Matrix} is input-to-state stable.
Thus, we get
\begin{align}
\|\tilde{{\delta}}_{m\tau,(i)}\| \leq \phi_1(\|{Z}_{\tau,(i)}\|)+\phi_2(\|\bar{{U}}_{l\tau,(i)}\|).
\end{align}

Since $\tilde{{\delta}}_{m\tau,(1)}=\mathbf{0}_{2n}$, we can obtain
\begin{align}\label{eqn:formation error norm}
\nonumber
\|\tilde{{\delta}}_{m\tau}\|&=\sqrt{\sum_{i=2}^{N}\|\tilde{{\delta}}_{m\tau,(i)}\|^2},\\
& \leq \sqrt{\sum_{i=2}^{N}(\phi_1(\|{Z}_{\tau,(i)}\|)+\phi_2(\|\bar{{U}}_{l\tau,(i)}\|))^2}.
\end{align}
Notice that the right-hand side of~\eqref{eqn:formation error norm} is independent of time instant, therefore, $\|\tilde{{\delta}}_{m\tau}\|$ is uniformly bounded.
Since~$\|\tilde{{\delta}}_{m\tau}\|~(m \in \mathbb{Z}_+)$ is bounded, based on~\eqref{eqn:Linear Transformation of Kronecker Basis},~$\|\tilde{{\delta}}_{m\tau+1}\|,\ldots,\|\tilde{{\delta}}_{(m+1)\tau-1}\|$ are all bounded, which indicates~$\|\tilde{{\delta}}_{k}\|$ for~$k \geq \tau$ is bounded.%

With ${\delta}_k=(Q \otimes I_{2n})\tilde{{\delta}}_k$ and $\|Q \otimes I_{2n}\|=1$, we have $\|{\delta}_k\| \leq \|\tilde{{\delta}}_k\|$.
Thus, the formation error $\|{\delta}_k\|$ is bounded.\hfill$\blacksquare$

\section{Proof of \thmref{thm:Connectivity preservation of control topology}}\label{apx:Proof of thm:Connectivity preservation of control topology}
Since the conditions of successful decoding~\eqref{eqn:conditions of successfully received-with uncertainty} is required to be satisfied from the beginning time instant of each message transmission, we need to predict the position ranges for each transmitter agent and their receiver agents during transmission.

For the control topology~$\mathcal{G}_c$, with the SNR condition~\eqref{eqn:SNR Condition}, if the transmit power of agent $i$ satisfies
\begin{equation}\label{eqn:power control}
P_{i,k}^{\mathrm{tx}} = \frac{(2^{\frac{\mu}{B_w}}-1)W_{j,k}}{g_{d_0}}(\frac{\bar{R}_{i,m\tau}}{d_0})^{\psi},~m=\lfloor \frac{k}{\tau_p} \rfloor,
\end{equation}
where~$\bar{R}_{i,m\tau}=\max_{j \in \mathcal{N}_{i,c}}\max_{l \in \{ 0,\ldots, \tau-1\}}{\tilde{R}_{i,m\tau+l}}$,
\begin{align}
\nonumber
\tilde{R}_{i,m\tau+l}=\max_{\substack{p_{i,m\tau+l} \in \llbracket \mathbf{p}_{i,m\tau+l} \rrbracket^{[i]} \\ \substack{p_{j,m\tau+l} \in \llbracket \mathbf{p}_{j,m\tau+l} \rrbracket^{[i]}}}}{\|p_{i,m\tau+l}-p_{j,m\tau+l}\|},
\end{align}
the decoding condition given in~\eqref{eqn:conditions of successfully received-with uncertainty} holds.
For the~$n$-dimensional balls~$\llbracket \mathbf{p}_{i,k} \rrbracket^{[i]}=\mathcal{B}[c_p(i,k),r_p(i,k)]$ and~$\llbracket \mathbf{p}_{j,k} \rrbracket^{[i]}=\mathcal{B}[c_p(j,k),r_p(j,k)]$, we can obtain
\begin{equation}
	\nonumber
	\tilde{R}_{i,k}=\|c_p(i,k)-c_p(j,k)\|+r_p(i,k)+r_p(j,k).
\end{equation}
In what follows, we give the explicit expression of $\tilde{R}_{i,k}$.

With \eqref{eqn:Modified Control Protocol}, the control inputs of each agent are predictable.
From the side of agent~$i$ at time~$k=m\tau~(m \in \mathbb{Z}_+)$, for agent~$i$, the predicted control input is:
\begin{equation}\label{eqn:Controller state estimation for I}
u_{i,m\tau+l}=KA^{\tau+l}\!\!\sum_{j \in \mathcal{N}_{i,c}}\!\!o_{ij}a_{ij}(x_{j,(m-1)\tau}-x_{i,(m-1)\tau}),
\end{equation}
where~$l=\{ 0,\ldots,\tau-2 \}$.
Since the available state information of agent~$j \in \mathcal{N}_{i,c}$ is~$x_{j,(m-1)\tau}$, it is necessary to predict the control inputs for agent~$j$ at time~$k \in [(m-1)\tau,(m+1)\tau-2]$.
For~$l \in \{ 0,\ldots,\tau-1 \}$, we have
\begin{equation}
\!\!\!\!u_{j,(m-1)\tau+l}=KA^{l}\!\!\!\!\sum_{j_1 \in \mathcal{N}_{j,c}}\!\!\!\!o_{jj_1}a_{jj_1}(\hat{x}_{j_1,(m-1)\tau}-\hat{x}_{j,(m-1)\tau}),
\end{equation}
where~$\sum_{j_1 \in \mathcal{N}_{j,c}}o_{jj_1}a_{jj_1}(\hat{x}_{j_1,(m-1)\tau}-\hat{x}_{j,(m-1)\tau})$ is acquired from messages transmitted by agent $j$, defined in~\eqref{eqn:Message Content}; for~$l \in \{ 0,\ldots,\tau-2 \}$, we have
\begin{equation}\label{eqn:non-accurate control of j}
\!\!u_{j,m\tau+l}=KA^{\tau+l}\!\!\!\!\sum_{j_1 \in \mathcal{N}_{j,c}}\!\!\!\!o_{jj_1}a_{jj_1}(\hat{x}_{j_1,(m-1)\tau}-\hat{x}_{j,(m-1)\tau}).
\end{equation}

Therefore, the position range of agent~$i$ at time~$k \in [m\tau+1,(m+1)\tau-1]$ is
\begin{align}\label{eqn:Prediction of I}
\nonumber
&\llbracket\mathbf{p}_{i,k}\rrbracket^{[i]}=\{p_{i,m\tau}\} \oplus (k-m\tau) h\{v_{i,m\tau}\} \\
\nonumber
&\oplus \sum_{l=1}^{k-m\tau}\llbracket\mathbf{w}_{i,m\tau+l-1}^{p}\rrbracket \oplus h\sum_{l=1}^{k-m\tau} (k-m\tau-l)\llbracket\mathbf{w}_{i,m\tau+l-1}^{v}\rrbracket \\
& \oplus h^2\sum_{l=1}^{k-m\tau}(k-m\tau+\frac{1}{2}-l)\{ u_{i,m\tau+l-1} \}.
\end{align}
The position range of agent~$j$ for~$k\in [m\tau,(m+1)\tau-1]$ is
\begin{align}\label{eqn:Prediction of J}
\nonumber
&\llbracket\mathbf{p}_{j,k}\rrbracket^{[i]}=\{p_{j,(m-1)\tau}\} \oplus (k-(m-1)\tau) h\{v_{j,(m-1)\tau}\} \\
\nonumber
&\oplus h^2\sum_{l=1}^{k-(m-1)\tau}(k-(m-1)\tau+\frac{1}{2}-l)\{ u_{j,(m-1)\tau+l-1} \} \\
\nonumber
&\oplus h\sum_{l=1}^{k-(m-1)\tau} (k-(m-1)\tau-l)\llbracket\mathbf{w}_{j,(m-1)\tau+l-1}^{v}\rrbracket \\
&\oplus \sum_{l=1}^{k-(m-1)\tau}\llbracket\mathbf{w}_{j,(m-1)\tau+l-1}^{p}\rrbracket.
\end{align}

Since the initial positions of all agents are known, the position range for agent $i \in \mathcal{V}$ at time $k \in [1,\tau-1]$ is
\begin{align}
\nonumber
\llbracket\mathbf{p}_{i,k}\rrbracket^{[i]}&=\{p_{i,0}\} \oplus k h\{v_{i,0}\} \oplus h^2\sum_{l=1}^{k}(k+\frac{1}{2}-l)\{ u_{i,l-1} \} \\
&\oplus \sum_{l=1}^{k}\llbracket\mathbf{w}_{i,l-1}^{p}\rrbracket \oplus h\sum_{l=1}^{k} (k-l)\llbracket\mathbf{w}_{i,l-1}^{v}\rrbracket,
\end{align}
where $u_{i,l-1}=KA^{l-1}\sum_{j \in \mathcal{N}_{i,c}}{a_{ij}(x_{j,0}-x_{i,0})}$.

Consequently, based on \propref{prop:Estimation of Uncertain Position Range}, the center and the radius of~$\llbracket \mathbf{p}_{s,k} \rrbracket^{[i]}=\mathcal{B}[c_p(s,k),r_p(s,k)]$,~$s \in \mathcal{N}_{i,c}\cup \{i\}$ are
\begin{align}
    \nonumber
	c_p(s,k)&=p_{s,k-\sigma}+\sigma hv_{s,k-\sigma}+h^2\sum_{l=1}^{\sigma} (\sigma+\frac{1}{2}-l)u_{s,k-\sigma+l-1}, \\
    \nonumber
    r_p(s,k)&=\sigma r_{w_p}(s)+\frac{1}{2}\sigma(\sigma-1)hr_{w_v}(s),
\end{align}
for~$\sigma>0$, and~$c(s,k)=p_{s,k},~r(s,k)=0$ for~$\sigma=0$.
With the above analysis, we conclude:~(i)~for agent~$i$ at time~$k \geq \tau$,~$\sigma=k-m\tau$;~(ii) for agent~$j$ at time~$k \geq \tau$,~$\sigma=k-(m-1)\tau$;~(iii) for agents~$i$ and~$j$ at time~$0 \leq k < \tau_p$,~$\sigma=k$.

Note that the actual control input of agent~$j$ at time~$k \in [m\tau,(m+1)\tau-2]$ should be 
\begin{equation}\label{eqn:accurate control of j}
\!\!\!\!u_{j,m\tau+l}=KA^{\tau+l}\!\!\!\!\sum_{j_1 \in \mathcal{N}_{j,c}}\!\!\!\!o_{jj_1}a_{jj_1}({x}_{j_1,(m-1)\tau}-{x}_{j,(m-1)\tau}),
\end{equation}
rather than that in~\eqref{eqn:non-accurate control of j}.
However, agent~$i$ cannot obtain the accurate state~${x}_{j_1,(m-1)\tau}~(j_1 \in \mathcal{N}_{j,c})$, which results in a deviation between the transmit power calculated by~\eqref{eqn:power control} based on~\eqref{eqn:non-accurate control of j} and~\eqref{eqn:accurate control of j} (the corresponding transmit power can be denoted by~$\bar{P}_{i,k}^{\mathrm{tx}}$ and~$\tilde{P}_{i,k}^{\mathrm{tx}}$).
To compensate for the power deviation, we modified the power condition~\eqref{eqn:power control} as follows
\begin{equation}
P_{i,k}^{\mathrm{tx}} = \frac{(2^{\frac{\mu}{B_w}}-1)W}{g_{d_0}}(\frac{\bar{R}_{i,m\tau}}{d_0})^{\psi} + \epsilon,
\end{equation}
where~$0 \leq \epsilon \leq \|\tilde{P}_{i,k}^{\mathrm{tx}}-\bar{P}_{i,k}^{\mathrm{tx}}\|$.\hfill$\blacksquare$

\bibliographystyle{IEEEtran}

\bibliography{Ref}

\vskip -2\baselineskip plus -1fil

\begin{IEEEbiography}[{\includegraphics[width=1in,height=1.25in,clip,keepaspectratio]{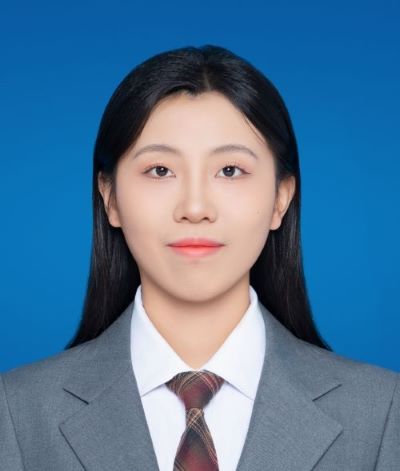}}]{Yaru Chen}
	 received the B.E. degree in automation from Northeastern University, Shenyang, China, in 2021. She is currently pursuing a Ph.D. degree in control science and engineering with the College of Intelligence Science and Technology at National University of Defense Technology, Changsha, China. Her research interests include cooperative control and wireless communications.
\end{IEEEbiography}

\vskip -2\baselineskip plus -1fil

\begin{IEEEbiography}[{\includegraphics[width=1in,height=1.25in,clip,keepaspectratio]{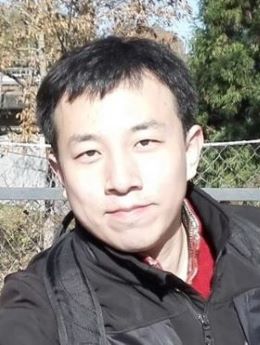}}]{Yirui Cong}
	(Member, IEEE) is an associate professor with National University of Defense Technology, Changsha, China. He received the B.E. degree (outstanding graduates) in automation from Northeastern University, Shenyang, China, in 2011, the M.Sc. degree (graduated in advance) in control science and engineering from National University of Defense Technology, Changsha, China, in 2013, and the Ph.D. degree from the Australian National University, Canberra, Australia, in 2018. His research interests include control theory, communication theory, and filtering theory.
\end{IEEEbiography}

\vskip -2\baselineskip plus -1fil

\begin{IEEEbiography}[{\includegraphics[width=1in,height=1.25in,clip,keepaspectratio]{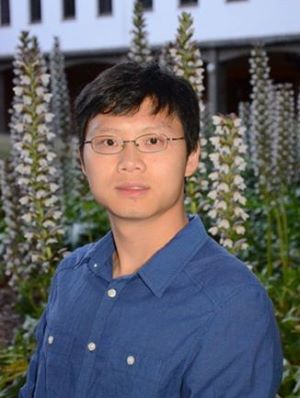}}]{Xiangyun Zhou}
	(Fellow, IEEE) is an Associate Professor at the Australian National University (ANU). He received the Ph.D. degree from ANU in 2010. His research interests are in the fields of communication theory and wireless networks. He has served as an Editor of IEEE TRANSACTIONS ON COMMUNICATIONS, IEEE TRANSACTIONS ON WIRELESS COMMUNICATIONS and IEEE WIRELESS COMMUNICATIONS LETTERS, and as an Executive Editor of IEEE COMMUNICATIONS LETTERS. He also served as symposium/track and workshop chairs for major IEEE conferences. He is a recipient of the Best Paper Award at ICC'11, GLOBECOM'22, ICC’24 and IEEE ComSoc Asia-Pacific Outstanding Paper Award in 2016. He was named the Best Young Researcher in the Asia-Pacific Region in 2017 by IEEE ComSoc Asia-Pacific Board. He is a Fellow of the IEEE.
\end{IEEEbiography}

\vskip -2\baselineskip plus -1fil

\begin{IEEEbiography}[{\includegraphics[width=1in,height=1.25in,clip,keepaspectratio]{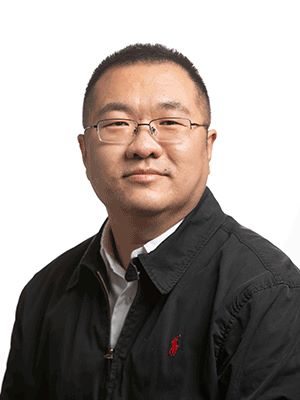}}]{Long Cheng}
	(Fellow, IEEE) received the B.S. degree (Hons.) in control engineering from Nankai University, Tianjin, China, in 2004, and the Ph.D. degree (Hons.) in control theory and control engineering from the Institute of Automation, Chinese Academy of Sciences, Beijing, China, in 2009. He is currently a Professor with the State Key Laboratory of Multimodal Artificial Intelligence Systems, Institute of Automation, Chinese Academy of Sciences. He is also a Professor with the University of Chinese Academy of Sciences, Beijing. He has published more than 100 technical papers in peer-refereed journals and prestigious conference proceedings. His research interests include rehabilitation robot, tactile sensor, human-robot interaction, intelligent control, and neural networks. Dr. Cheng was a recipient of IEEE Transactions on Neural Networks Outstanding Paper Award from the IEEE Computational Intelligence Society, the Aharon Katzir Young Investigator Award from the International Neural Networks Society, and the Young Researcher Award from the Asian Pacific Neural Networks Society. He is an Associate Editor of IEEE Transactions on Cybernetics, IEEE Transactions on Automation Science and Engineering, IEEE Transactions on Cognitive and Developmental Systems, IEEE/ASME Transactions on Mechatronics, Science China Information Sciences, and Acta Automatica Sinica. He is the IEEE Fellow and IET Fellow.
\end{IEEEbiography}

\vskip -2\baselineskip plus -1fil

\begin{IEEEbiography}[{\includegraphics[width=1in,height=1.25in,clip,keepaspectratio]{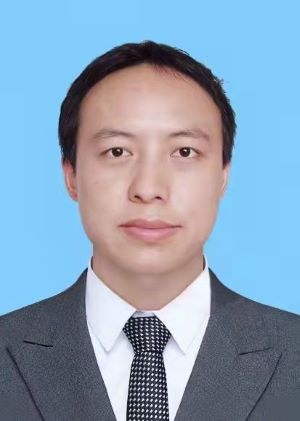}}]{Xiangke Wang}
	(Senior Member, IEEE) received the B.S., M.S., and Ph.D. degrees in Control Science and Engineering from National University of Defense Technology, China, in 2004, 2006, and 2012, respectively. Since 2012, he has been with the College of Intelligence Science and Technology, National University of Defense Technology, where he is currently a full professor. He was a visiting student at the Research School of Engineering, Australian National University from 2009 to 2011. He is a senior member of IEEE and is supported by the Hunan Outstanding Youth Award Program. His current research interests include the control of multi-agent systems and its applications on unmanned aerial vehicles. He has authored or coauthored 5 books and more than 200 papers in refereed journals or international conferences, including IEEE Transactions/Letters, CDC, IFAC, ICRA. etc.
\end{IEEEbiography}

\end{document}